\documentclass[9pt]{elsarticle}
\usepackage[utf8]{inputenc}
\usepackage{graphicx}
\usepackage{amsmath}
\usepackage[margin=1.2in]{geometry}

\usepackage{sectsty}
\usepackage{float}
\usepackage{gensymb}
\usepackage{amssymb}

\newcommand{\new}[1]{{#1}} 
\newcommand{\newc}[1]{{#1}} 
\newcommand{\newd}[1]{{\index{ignore}}} 

\title{Quasi-universal scaling in mouse-brain neuronal activity stems from edge-of-instability critical dynamics}
\author[a]{Guillermo B. Morales}
\author[b]{Serena Di Santo}
\author[a]{Miguel A. Mu{\~n}oz \corref{cor1}}
\cortext[cor1]{Corresponding author mamunoz@onsager.ugr.es}

\address[a]{Departamento de Electromagnetismo y F{\'\i}sica de la
  Materia and Instituto Carlos I de F{\'\i}sica Te{\'o}rica y
  Computacional. Universidad de Granada.  E-18071, Granada, Spain}
\address[b]{Center for Theoretical Neuroscience, M.
B. Zuckerman Mind Brain Behavior Institute, Columbia University, New York, NY USA}

\begin{document}
\begin{abstract} The brain is in a state of perpetual reverberant neural activity, even in the absence of specific tasks or stimuli. Shedding light on the origin and functional significance of \new{such a dynamical state} is essential to understanding how the brain transmits, processes, and stores information. An inspiring, albeit controversial, conjecture proposes that some statistical characteristics of empirically observed neuronal activity can be understood by assuming that brain networks operate in a dynamical regime near the edge of a phase transition. Moreover, the resulting critical behavior, with its concomitant scale invariance, is assumed to carry crucial functional advantages. Here, we present a data-driven analysis based on simultaneous high-throughput recordings of the activity of thousands of individual neurons in various regions of the mouse brain. To analyze these data, we \newd{construct a unified theoretical framework that} synergistically combine\newd{s} cutting-edge methods for the study of brain activity (such as a phenomenological renormalization group approach and techniques that infer the general dynamical state of a neural population), while designing complementary tools. This strategy allows us to uncover strong signatures of scale invariance that is "quasi-universal" across brain regions and reveal that all these areas operate, to a greater or lesser extent, near the edge of instability. Furthermore, this framework allows us to distinguish between quasi-universal background activity and non-universal input-related activity. Taken together, this study provides strong evidence that brain networks actually operate in a critical regime which, among other functional advantages, provides them with a scale-invariant substrate of \new{activity covariances that can sustain optimal input representations}. \end{abstract}




\maketitle

\section{Introduction}
The  \newd{cerebral cortex} \new{brain} of mammals is in a state of continuous ongoing activity even in the absence of stimuli or specific tasks \cite{Softky,Arieli,Raichle}. Shedding light onto the origin and functional meaning of such an energy-demanding baseline state of background dynamics and its interplay with input-evoked activity are challenging goals, essential to ultimately understand how \newd{the cortex} \new{different brain regions} represent, process and transmit information \cite{Friston,Stringer-Science,Stringer-Nature}.  Following the fast-paced development of powerful neuroimaging and electrophysiological technologies such as two-photon calcium imaging 
and neuropixels probes, 
recent years have witnessed important advances in our understanding of these issues.

An inspiring hypothesis ---which aims to become a general principle of brain dynamical organization--- posits that neuronal networks could achieve crucial functional advantages, including optimal information processing and transmission, by operating in the vicinity of a critical point \cite{BP2003,Mora-Bialek,Bialek-book,Chialvo2010,RMP}. Using the jargon of statistical physics, this implies that the network operates in an intermediate regime at the border between "ordered" and "disordered" phases \cite{BP2003,Chialvo2010,Plenz-functional,Plenz-task,Breakspear-review,Lucilla,Viola-review,Plenz-review,Levina,Martinello,Fosque21,LG}.  Criticality, with its concomitant power-laws, entails well-known functional advantages for information processing such as, e.g., an exquisite sensitivity to perturbations and a huge dynamic range \cite{Plenz-functional,RMP} to name a few.  \new{Moreover}, it also gives rise to the emergence of a broad spectrum of spatio-temporal scales, i.e. scale invariance or simply "scaling" \cite{RMP,Newman,Binney}, that could be essential to sustain brain activity at a broad hierarchy of scales, as required to embed highly-diverse cognitive processes.

In spite of its conceptual appeal and thrilling implications, the validity of this so-called "criticality hypothesis" as an overarching principle of dynamical brain organization remains controversial \cite{Destexhe,Beggs-critical,Marsili-2019}. 
Therefore, novel  theoretical approaches and more-stringent experimental tests are much needed to either prove or disprove this conjecture and, more in general, to advance our understanding of the mechanisms underlying brain ongoing activity.

From the theoretical side it is crucial to refine the conjecture itself and discern what type of criticality is the most pertinent to describe brain activity \cite{Mora-Bialek,RMP}.  Different possible scenarios have been explored; among them \cite{RMP}:  (i) the edge of
activity propagation, with scale-free avalanching behavior \cite{BP2003,Chialvo2010,Breakspear-review,Moretti}, (ii) the edge of a synchronization  phase transition
\cite{Plenz-synchro1,LG,Zhou-Hopf,Villegas,hybrid,RMP}, and (iii) the edge of a stable-unstable transition, often called "edge of chaos" but to which we will refer as "edge of instability" hereon \cite{Steyn,Magnasco,Helias}.

From the empirical side, evidence of putative brain criticality often relies on the detection of scale-free bursts of activity, called "neuronal avalanches"  \cite{BP2003,Petermann,Plenz-review,Turrigiano,Zhou-Hopf,Fontenele}, the presence of robust long-range spatio-temporal correlations \cite{Linken-2012,Linken-2013}, the analysis of large-scale whole-brain models fitted to match empirically observed correlations  \cite{Cabral-review,Taglia,Zhou-jointly,Zhou22}, etc.
Recently, with the advent of modern techniques enabling simultaneous recordings of thousands of neurons,
complementary experimental evidence revealing the existence of scaling in brain activity ---which might or might not stem from underlying criticality  \cite{Mora-Bialek,RMP,Marsili-2019,Friston-scale}--- 
has emerged from unexpected angles. 
Among these, let us mention (i) a novel renormalization-group approach which identified strong signatures of scale-invariant activity in recordings of more than a thousand neurons in the mouse hippocampus \cite{Bialek-PRL,Bialek-largo}, (ii) the direct inference ---using linear response theory and methods from the physics of disordered systems--- of "edge-of-instability" type of critical behavior in neural recordings from  the macaque monkey motor cortex \cite{Helias}, and (iii) the discovery of an unexpected \new{power-law distribution, which revealed scale invariance in the spectrum of the covariance-matrix from} tens of thousands of neurons in the mouse visual cortex 
while the mouse was exposed to a large set of sequentially-presented natural images
\cite{Stringer-Nature}.

What all these works have in common, is the fact the they use specifically-designed, powerful mathematical tools to analyze vast amounts of high-throughput data.
\newd{Thus, these approaches not only open up new and exciting research avenues but, even more importantly, they allow for synergistic approaches to exploit them together, paving the road to further advances.}

Here we \new{bring these diverse approaches to a common ground and develop novel and complementary tools}
\newd{propose an unified theoretical framework and employ it} \new{leveraging them synergistically} in order to analyze state-of-the-art neural recordings in diverse areas of the mouse brain \cite{Steinmetz}. As shown in what follows, these analyses strongly enhance our understanding of scale-invariance and possible criticality in the brain. In particular, they allow us to elucidate the emergence of quasi-universal scaling across regions in the mouse brain,  to conclude that all such regions are, to a greater or lesser extent, posed close to the edge of instability, as well as to disentangle background and input-evoked neural activities. 

\section*{Theoretical framework and open-ended questions}

For the sake of self-containedness, let us briefly discuss the three innovative theoretical approaches cited above, along with some stemming open-ended questions that we pose (further technical details are deferred to the Methods section).

\vspace{0.2cm}

(A) \textbf{Renormalization-group approach to neuronal activity.} 
The renormalization group (RG) is retained as one of the most powerful ideas in theoretical physics, allowing us to rationalize collective behavior ---at broadly-diverse observational scales--- from the properties of the underlying "microscopic" components,  and to understand, for instance, the emergence of scale invariance \cite{Binney}. In a remarkable contribution, Meshulam {\emph et al.} developed a\newd{n} \new{phenomenological} RG approach to analyze  time series from  large populations of simultaneously recorded individual spiking neurons and scrutinize their collective behavior \cite{Bialek-PRL,Bialek-largo}. The method, similar in spirit to Kadanoff's blocks for spin systems \cite{Binney}, allows one to construct effective descriptions of time-dependent neural activity at progressively larger "coarse-grained" scales. Notably, a number of non-trivial features ---generally attributed to scale-invariant critical systems---  emerge from the application of such RG analyses to recordings of more than $1000$ neurons in the mouse hippocampus while it is moving in a virtual-reality environment. These features include among others: (i) a non-Gaussian (fixed-point) distribution of neural activity at large scales, (ii) non-trivial scaling of the activity variance and auto-correlation time as a function of the coarse-graining scale, and (iii) a power-law decay of the spectrum of the covariance matrix \cite{Bialek-PRL,Bialek-largo,Bradde}. 

A limitation of this type of phenomenological RG analysis is that, even if it is capable of uncovering scale invariance, it does not allow discerning what kind of putative phase transition could be at \new{its origin}. \newd{roots of the observed scale.} Moreover, doubts have been raised about the possible interpretation of the results as stemming from criticality
(see \cite{Nemenman-PRL,Nicoletti} and below).

In any case, leaving aside for the time being these caveats, one can wonder whether the observed scaling features are shared by other brain regions or if they are rather specific to the mouse hippocampus. Is there any kind of "universality \new{or at least "quasi-universality" (in the sense of scaling exponents showing limited variability)} in the neural dynamics across brain regions despite their considerable anatomical and functional differences? Is it possible to find empirical evidence that allows us to  discern whether the observed scaling actually stems from critical behavior\newd{and is not a mere consequence of, for example, unobserved latent inputs}? 

\vspace{0.2cm}

(B) \textbf{Inferring the dynamical regime from neural recordings.}
Dahmen \emph{et al.} devised a general approach ---based on linear-response theory ideas and tools from the physics of disordered systems--- that allows one to infer the overall dynamical state of an empirically observed neuronal population. In particular, this theoretical approach permits us to \new{estimate} the distance to the "edge of instability" from empirical measurements of the mean and dispersion of "spike-count covariances" (also called "long-time-window" covariances) across pairs of recorded neurons \cite{Helias}. Straightforward application of this approach to neural recordings from motor cortex of awake macaque monkeys \new{at rest} strongly supports the idea that such a region operates in a "dynamically balanced critical regime with nearly unstable dynamics" \newd{, that they called "type-II criticality"  (see below)} \cite{Helias}. Thus, one can wonder whether  other brain regions ---for instance, the hippocampus in (A)--- can also be empirically proven to be similarly close to the edge of instability.

Moreover, recently  Hu and Sompolinsky  went a step further and derived  an analytical expression \new{(based also on linear-response theory for large random networks)} for the full spectrum of eigenvalues of the spike-count covariance matrix as a function of its distance to the edge of instability.
\new{This provides us with an alternative method to estimate the distance to the edge of instability from empirical data.} In particular, these analyses reveal that the eigenvalue distribution develops a power-law tail as \newd{(and only if)} the network approaches the edge of instability \cite{YuHu}.  The resulting  non-trivial eigenvalue distribution stems from the recurrent network dynamics near the edge of instability \cite{YuHu} \new{and it clearly differs from the Marchenko-Pastur law  for the correlations of independent random units} \newd{,  so that it cannot be possibly ascribed to, e.g., limited sampling} \cite{cleaning}. 
\newd{or hidden variables \cite{Nemenman-PRL}.}
Thus, we also infer what is the distance to the edge-of-instability in the theoretical model that best fit the spectra of spike-count covariance matrices obtained from empirical data.

\newpage

(C) \textbf{Scaling in optimal input representations.} 
Stringer \emph{et al.} \cite{Stringer-Nature} studied  \new{experimentally and theoretically} the spectrum of the covariance matrix in neuronal populations in \newd{the} mouse \newd{V1} \new{primary} visual cortex \new{(VISp)} while the mouse was exposed to a very large set  of sequentially-presented natural images (recording more than $10^4$ neurons in parallel; see Methods). From the  resulting data they found that the spectrum of the covariance matrix obeyed  "\emph{an unexpected power law}" \cite{Stringer-Nature}:  the n-th rank-ordered eigenvalue scaled as $1/n^{\mu}$ with $\mu \gtrsim 1$. 

This power-law decay of the rank-ordered eigenspectrum was somehow surprising; the authors were expecting a much faster decay, as would correspond to a lower-dimensional representation of the visual inputs (see below). Notice that, here, by "representations" one means neural activity that stems from or is correlated with sensory or task-related inputs. On the other hand, the term "dimensionality" is employed in the sense of "principal component analysis" (PCA) \cite{PCA}, where 
the dimension is the number of principal components required to explain a given percentage of the total variance; often  most of the variability in neural data can be recapitulated in just a few principal components or dimensions \cite{Ganguli}, \new{but this is not always the case \cite{Stringer-Nature}.}

Remarkably, Stringer \emph{et al.} were also able to prove that the power-law decay of the rank-ordered eigenvalues is not an artifact directly inherited from the statistics of the inputs, but a very deep mathematical property: it stems from a trade-off between the neural representation of visual inputs  being as high-dimensional as possible (i.e. including a large number of non-negligible components in a PCA analysis) and mathematically preserving its smoothness (i.e. its continuity and differentiability). As a simple illustration of this last abstract property, let us mention that the smoothness of the representation prevents, for instance, that tiny variations in the inputs dramatically alter the neural population activity, which translates into \new{a more robust encoding} \cite{Stringer-Nature}. 

Let us finally recall that common knowledge in statistical physics tells us that a power-law decay of the covariance-matrix spectrum (i.e., of the "\emph{propagator}") is one of the most remarkable generic trademarks of critical behavior, emphasizing the emergence of a scale-free hierarchical organization of spatio-temporal correlations \cite{Binney}. Indeed, this spectrum is one of the objects studied in the phenomenological RG approach (A), revealing a power-law decay (for the mouse hippocampus) with an exponent $\mu <1$ \cite{Bialek-PRL}, which \newd{disagrees with} \new{violates} Stringer \emph{ et al.}'s \newd{predictions} \new{bound for continuous neural representations} ($\mu \gtrsim 1$).

The results by Stringer \emph{et al.} trigger a cascade of questions: is the empirically-observed scaling of the spectrum of the covariance matrix a \new{mere} consequence of the external input being represented in an optimal way? In other words, does the covariance spectrum obey scaling also in the absence of inputs, i.e. for resting-state or ongoing activity? Is intrinsic criticality in the network dynamics required to "excite" such a broad spectrum of modes supporting optimal input representations? How come the exponent $\mu$ \new{measured by Meshulam et al.} in the hippocampus is smaller than $1$ in seeming contradiction with Stringer \emph{et al.}'s predictions?

\vspace{0.5cm}

\new{Summing up:}  while (A) allows us to detect and scrutinize the presence of scaling in empirical recordings in a systematic and quantitative way,  (B) provides us with practical tools to infer the actual dynamical regime of the underlying neural network,  thus paving the way to ascribe empirically-reported scaling to edge-of-instability criticality, and  (C) \new{sets the scene for relating} \newd{paves the way to relate} criticality to so far unexplored functional advantages for optimal input representation.

In what follows, we use these methods in a synergetic way while developing novel and complementary tools to scrutinize scale invariance and criticality across regions in the mouse brain, thus providing data-driven answers to most of the previously raised questions. \newd{to create a unified theoretical framework which allows us}

 \section*{Results}
 
 \begin{figure}
 \begin{center}
\includegraphics[width=12 cm]{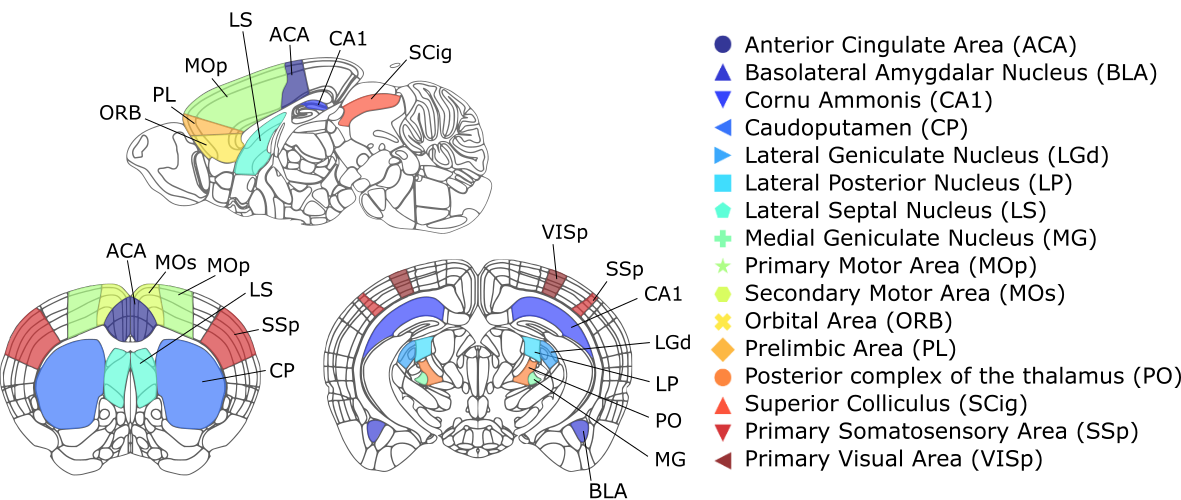}
\caption{Schematic representation of the regions in the mouse brain ---using three different projections--- considered in this work, together with their names and corresponding acronyms. Credit: Allen Institute, Atlas brain maps: https://atlas.brain-map.org/atlas.}\label{Brain_Regions}
\end{center}
\end{figure} 

\new{Most of the forthcoming analyses rely on} the empirical electrophysiological data presented by Steinmetz \emph{et al.} in \cite{Steinmetz},  where the activity, $x(t)$, of thousands of individuals neurons (in  particular, the precise times of their spikes) is simultaneously recorded at a high $200$Hz resolution in several mouse brain regions (as illustrated in Fig.\ref{Brain_Regions}).
These recordings include periods in which the mouse is performing some specific task and some in which is at a "resting-state".  Therefore, we first separate both types of time series and analyze 
the corresponding  "resting-state" \newd{(to which we also refer as "background" or "ongoing")} and "task-related" activity independently, paying special attention to the former. In addition, we also consider data from recordings of \new{mouse VISp} \newd{the V1 visual cortex} from Stringer {\emph et al.} \cite{Stringer-Nature} (see Methods). In all cases, we restrict our analyses to areas (see Fig.\ref{Brain_Regions}) with at least \new{$N=128$} simultaneously-recorded neurons. \newd{to ensure that scaling, if any, could be potentially observed for at least two decades.}

\subsection*{Quasi-universal scaling across brain regions}
We first employ the phenomenological RG approach to scrutinize whether non-trivial scaling behavior, such as the one reported in \cite{Bialek-PRL,Bialek-largo} for the \new{CA1 region of} mouse hippocampus, is observed in other areas of the mouse brain. For the sake of clarity let us summarize the gist of the RG approach (further details in the Methods Section as well as in \cite{Bialek-PRL, Bialek-largo}), along with our main results.

\begin{figure*}[!ht]
\centering
\includegraphics[width=14 cm]{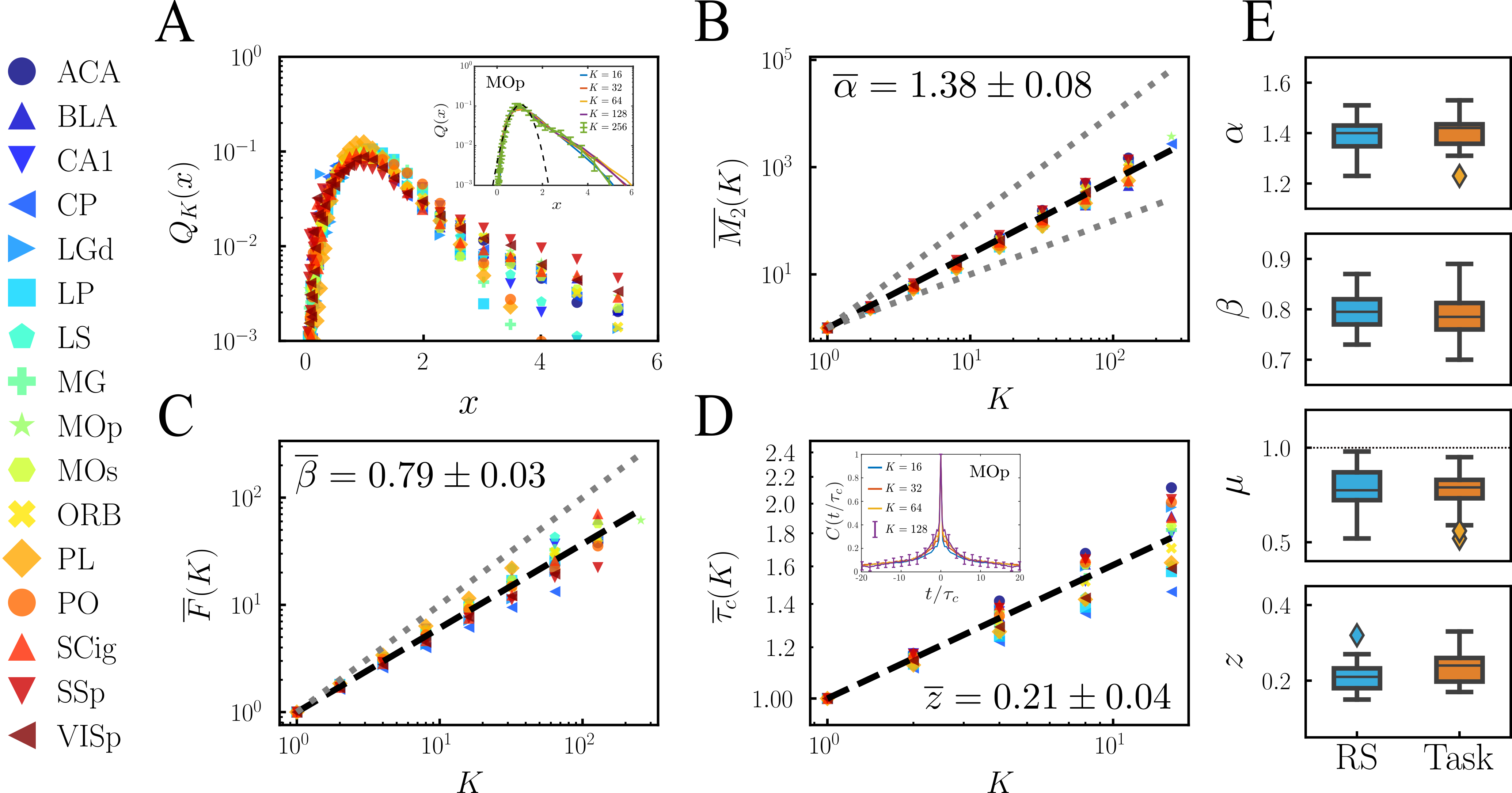}
\caption{Results of the phenomenological RG analyses of brain activity measured in \new{$16$} different mouse brain areas (A-D resting state activity). (\textbf{A}) Probability distribution for normalized non-zero activity in block-neurons of size $K=32$ across brain regions (main panel), as well as (inset) at $5$ consecutive steps of the coarse-graining  for a representative region (MOp). (\textbf{B}) Variance of the non-normalized activity as a function of the block-neuron size, $K$, in double logarithmic scale (upper and bottom dotted lines, with slopes $2$ and $1$, mark the fully correlated and independent limit cases, respectively). (\textbf{C}) Scaling of the free energy $F_{k}$
as defined in Methods (the dotted line corresponds to the expected behavior for un-correlated variables).
(\textbf{D}) Inset: decay of the autocorrelation function as a function of the rescaled time $t/\tau_c$ for the MOp region; a different value of $\tau_c$ is used for each cluster size but, after rescaling, all data collapse into a common curve.
Main plot:  scaling of the characteristic correlation time $\tau_c$ as a function of  $K$ in double logarithmic scale for the different areas. To facilitate the comparison between areas with different number of neurons, the variance has been normalized as $\overline{M_2}(K) = M_2(K)/M_2(K=1)$, the probability of being silent re-scaled to $\overline{F}(K) = F(K)/F(K=1)$ and the
correlation time as $\overline{\tau_{c}}(K) = \tau_{c}(K)/\tau_{c}(K=2)$. (\textbf{E}) Comparison of the exponent values $\alpha, \beta, \mu$ and $z$ for resting-state (RS) and task-related activity (Task). Horizontal line inside each box represents the sample median across regions, the whiskers reach the non-outlier maximum and minimum values, and the distance between the top (upper quartile) and bottom (lower quartile) edges of each box is the inter-quartile range (IQR). Outliers are represented with a diamond marker. For the exponent $\mu$, the critical value $\mu^{c} = 1$ has been marked with a dotted line.}
\label{4Measures}
\end{figure*}  
Following the spirit of Kadanoff's block one seeks to perform a coarse graining of $N$ "microscopic variables" ---i.e. single-neuron activities in this case--- to construct effective descriptions at progressively larger scales. Nevertheless, given the absence of detailed information about physical connections (synapses) between neurons, a criterion of maximal pairwise correlation (rather than the standard one of maximal proximity) is used to block together pairs of neurons in a sequential way \cite{Bialek-PRL}. Let us remark that, for this, it is crucial to determine pairwise correlations in a careful and consistent way; this requires the choice of a  suitable discrete time bin for each data set, for which we have devised an improved protocol (see \textit{Time-scale determination} in the Extended Methods of the SI). In this way, the activity time series of the two most correlated neurons are added together and properly normalized, giving rise to effective time series for "\emph{block-neurons}" or simply "\emph{clusters}" of size $2$. One then proceeds with the second most-correlated pair of neurons and so on, until all neurons have been grouped in pairs. The process is then iterated in a recursive way, so that after $k$ coarse-graining (RG) steps there remain only $N_{k}=N/2^{k}$ block-neurons, each recapitulating the activity of $K=2^{k}$ individual neurons. 
 
The distributions of activity values across block-neurons at level $k$, $P_k(\{x \})$, can be then directly computed for the different steps of the RG clustering procedure and, from them, a number of non-trivial features can be identified for all the considered mouse brain areas \cite{Steinmetz}:
 
\textbf{Non-Gaussian probability distribution of block-neuron activity}. Fig.\ref{4Measures}A shows the probability distribution $Q_{k}(x)$ of non-zero activity values, $x$,  of the coarse-grained block-neurons at one of the RG steps ($k=5$, i.e., $K=32$).  Observe also, in the inset of Fig.\ref{4Measures}A, the curve collapse obtained for sufficiently large values of $K$ as well as the presence of significant non-Gaussian tails, as exemplified  for one of the considered brain regions: the primary motor cortex \new{(MOp)}. This convergence of $Q_{k}(x)$ to an asymptotic shape after a few RG steps, as observed for all regions (see Fig.S1 in the SI), suggests that a nontrivial fixed point of the RG flow has been reached and that the emerging distribution of activity is rather universal across brain regions in resting conditions.

\textbf{Scaling of the activity variance}. As shown in Fig.\ref{4Measures}B an almost perfect scaling is observed for the variance $M_2(K)$ of the non-normalized activity of block neurons, with an average exponent \new{${\overline{\alpha}}=1.38 \pm 0.08$} across regions. In particular, we measure \new{${\alpha_{CA1}=1.37 \pm 0.03 \pm 0.02}$ (mean + mean-absolute-error of individual measurements + standard deviation over experiments, see Methods)} for the CA\new{1} region within the hippocampus, which is within errorbars of the value \new{${\alpha=1.56 \pm 0.07 \pm 0.16}$} \new{reported in} \cite{Bialek-largo} for \newd{such an }\new{the same} area. 
Notice that all these exponent values are always in between the expected ones for uncorrelated ($\alpha=1$) and fully-correlated variables ($\alpha=2$), revealing consistently the existence of non-trivial scale-invariant correlations. 

\textbf{Scaling of the "free-energy".} This is defined as ${F(K) = -\log(S_K)}$, where $S_K$ is the probability for a block-neuron \newd{at step $k$} of size $K$ to be silent \newd{(i.e. $x=0$)} within a time bin. As shown in Fig.\ref{4Measures}C,  this quantity exhibits a clear scaling with cluster size, with an average exponent \new{$\overline{\beta} = 0.79 \pm 0.03$ across regions ($0.78 \pm 0.04 \pm 0.05$ for CA1, to be compared with the value $0.87 \pm 0.014 \pm 0.015$} reported in \cite{Bialek-largo}). 

 \textbf{Scaling of the auto-correlation time} Fig.\ref{4Measures}D shows that, despite the very broad variability of intrinsic timescales for individual neurons within each region (see Fig.S3 in the SI), 
 dynamical scaling can be observed  in the decay of the block-neuron autocorrelation times $\tau_{c}(K)$ (see Methods) in all regions, with an average exponent across regions \new{$\overline{z}=0.21 \pm 0.04$ (with $z_{CA1}=0.18 \pm 0.03 \pm 0.01$  for \newd{the} CA\new{1} \newd{region in the hippocampus}, in perfect agreement with the one reported in \cite{Bialek-largo} for this region ($z=0.22 \pm 0.08 \pm 0.10$)} and compatible also with the 
 exponent values \new{reported in \cite{Friston-scale} using a different approach}). As an additional test for dynamical scaling, we show in the SI (Fig.S2) how the curves for the auto-correlation functions at different coarse-graining levels collapse when time is appropriately re-scaled.

\begin{figure}[ht!]
\begin{center}
\includegraphics[width=12cm]{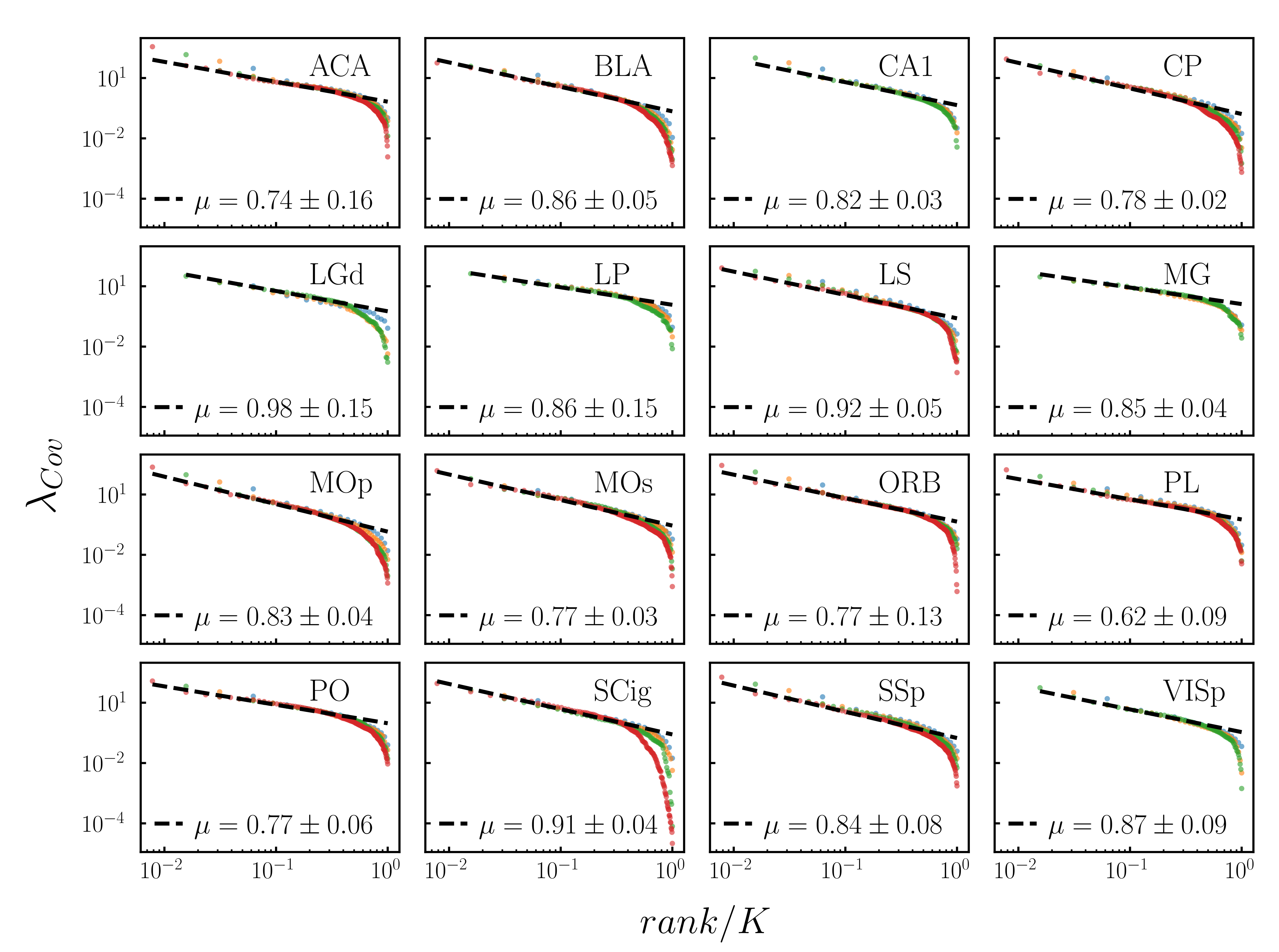}
\caption{Scaling of the covariance matrix spectrum for the resting-state activity in clusters of size $K=\{16, 32, 64, 128\}$ (blue, yellow, green and red markers, respectively) in \new{$16$} different brain regions. Observe not only the decay  of the rank-ordered eigenvalues  as a power law of the rank, but also the excellent collapse of the cut-offs obtained after rescaling the eigenvalue rank by the total size $K$.
For each region, we collect the average powerlaw exponent and its standard deviation across experiments (see also Table S2 in the SI). \label{Eigenvalues_Spont}}
\end{center}
\end{figure}

 \textbf{Scaling of the covariance-matrix spectrum.} By diagonalizing the covariance matrix computed at different levels of coarse-graining, it is possible to analyze  how their corresponding spectra decay with the rank of the eigenvalues and how their cut-offs change with cluster size. As illustrated in Fig.\ref{Eigenvalues_Spont}, in all the analyzed brain areas there is a clear power-law scaling of the eigenvalues with the rank, with an average exponent across regions \new{$\overline{\mu}=0.84 \pm 0.14$}, as well as a common dependence with the fractional rank ($rank/K$), the latter manifested in the collapse of the curves at different levels of coarse graining, much as in \cite{Bialek-PRL}. Likewise, \new{the value reported in \cite{Bialek-PRL} for \newd{the} CA1 \newd{region in the hippocampus} ($\mu = 0.76 \pm 0.05 \pm 0.06$) is in perfect agreement with our measured value for the same region, $\mu_{CA1} = 0.78 \pm 0.08 \pm 0.02$.} Although we will address this point later on, let us for now stress that $\mu$ is smaller or, at most, approximately equal to one $1$ in all regions, \new{seemingly suggesting discontinuity of the neural representations \cite{Stringer-Nature}}.

For the sake of consistency, we have also verified that the reported exponent values exhibit little variability upon changes in the time-discretization bin, with the exception of the exponent $\mu$, for which we observe an increase on longer time-scales beyond the typical inter-spike-interval of the population activity (see Fig.S5 in SI).

Moreover, as it turns out, similar signatures of scale-invariance to those reported for resting-state activity emerge in RG analyses of neural recordings obtained while the mice are performing a \newd{visual discrimination} task (see section \textit{Datasets} in Extended Methods of the SI and \cite{Steinmetz} for more details). This similarity is illustrated in Fig.\ref{4Measures}E, which shows how the dispersion and mean value of the scaling exponents across regions are not significantly altered \new{($p>0.1$ on a two-sample t-test for each exponent) when one compares the resting-state and task-related activity (see also Fig.S6).}

\new{Finally, as a control test we verified that the non-trivial scaling features revealed by the RG analyses are lost for all areas when the correlation structure of the data is broken either by: (i) reshuffling the times of individual spikes in the time series; (ii) shifting each individual time series by a random time span while keeping the sequence of spikes; or (iii) shuffling spikes across neurons (see Figs.S12-S14 in SI).}

\subsection*{Dynamical state of \newd{background} \new{resting-state} activity across brain regions}

Despite the elegance and appeal of the RG results ---as originally presented by Meshulam {\emph et al.}--- their straightforward interpretation as stemming from underlying criticality has been questioned \cite{Nemenman-PRL,Nicoletti}. In particular, very similar scaling behavior was found to emerge in a (non-critical) model of uncoupled neurons exposed to latent correlated inputs \cite{Nemenman-PRL} (see Discussion). Thus, it is not  guaranteed \emph{a priori} that the  scaling behavior we just found across brain regions stems from underlying criticality and the RG approach does not allow us to provide an answer to this question. Therefore,  we resort to alternative methods to estimate the dynamical regime of each brain region from empirical data
as described above \cite{Helias,YuHu}. For this, one needs to compute spike-count covariance values across pairs of neurons, which measure the pairwise correlations in time-integrated activity (i.e. the total number of spikes) across samples \newd{, thus revealing intrinsic features of the underlying network and dynamical regime} (see Methods).

\begin{figure*}[!ht]
\centering
\includegraphics[width=14cm]{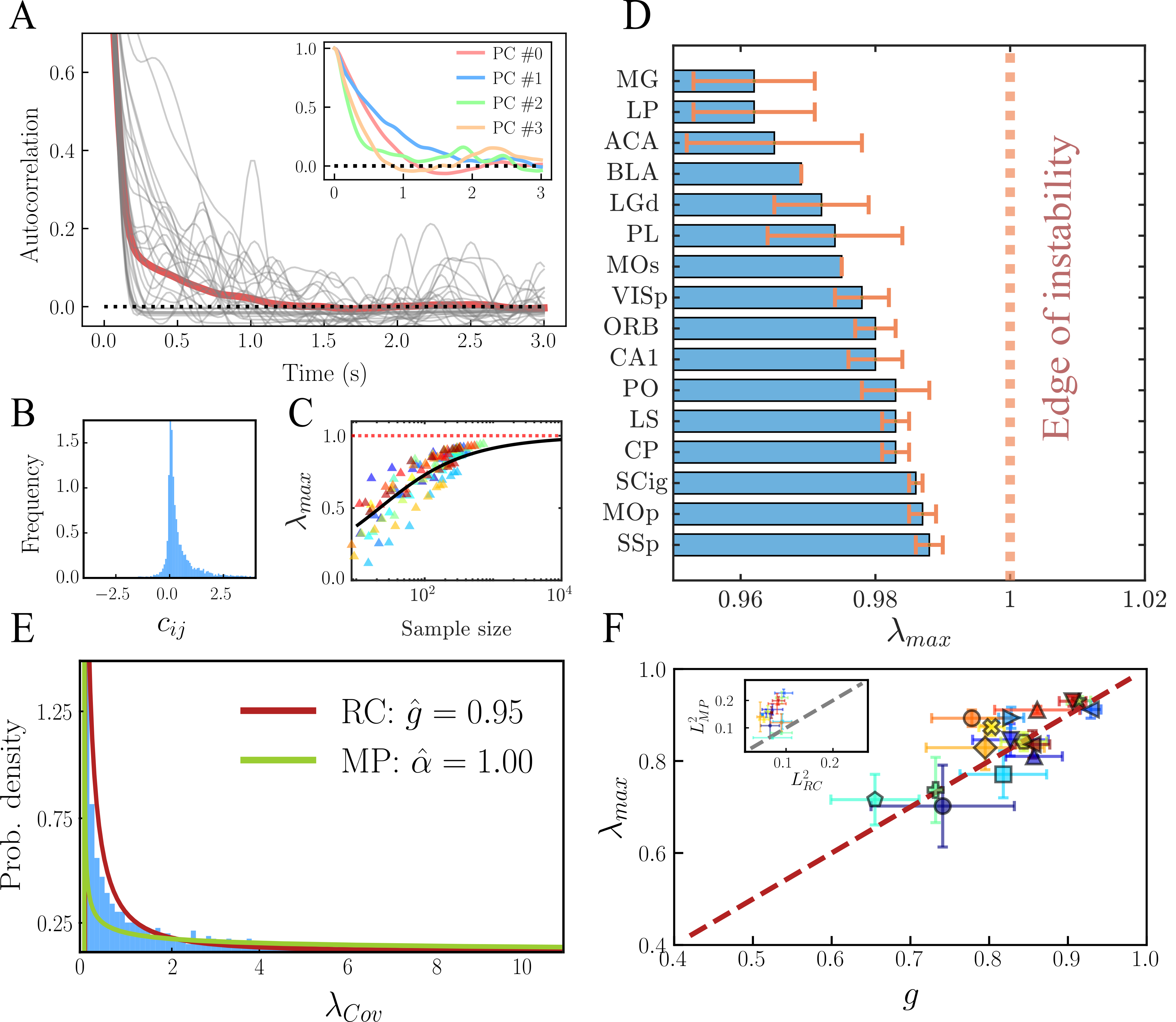}
\caption{(\textbf{A}) \new{Autocorrelations decay in the MOp region 
(grey lines correspond to $30$ randomly chosen neurons, while red line denotes the average across all neurons). Notice how autocorrelations vanish on a timescale of $\sim1s$. Inset:  Decay of the average autocorrelation for the firing rates projected into the first four principal components.}
(\textbf{B}) Distribution of pairwise covariance values in one of the considered brain regions (MOp). Observe the strong peak around $0$  (as expected from theoretical approaches of "balanced networks" \cite{Renart}) as well as the large dispersion of values \cite{Helias,YuHu}, including (asymmetric) broad tails. (\textbf{C}) Dependence of the estimated value of $\lambda_{max}$ on the number of neurons $N$ in the sample. Each color corresponds to a region following the color code in Fig.\ref{Brain_Regions}; for each region we show $10$ points, corresponding to sub-samples of $10 \%$ to $100 \%$ of the total neurons recorded, using the experiment with a greater number of recorded neurons for each area. Fitting the resulting values of $\lambda_{max}$ for different values of $N$ to Eq.\ref{lambda_max} with a parameter $\Delta$ common to all regions and extrapolating for large values of $N$, one observes a fast convergence to the very edge of instability. (\textbf{D}) Proximity to criticality in \new{$16$} different regions of the mouse brain, as measured by the estimated largest eigenvalue $\lambda_{max}$ ---with $\lambda_{max}=1$ marking the "edge of instability"--- of the inferred connectivity matrix.\newd{;the reported values can be considered \emph{lower bounds} of the actual maximal eigenvalues (see panel C).} \new{The regions are ordered according to their average $\lambda_{max}$ and the error bars are calculated as standard deviation over different experiments in the same region.}\newd{Rank-ordered eigenvalues of the spike-count covariance matrix for $13$ of the analyzed areas. In all cases, the decay is well described as a power law for at least two decades, with exponent values as reported in the legends, not far from  the theoretical prediction (for the case of Gaussian-distributed weights) in networks operating at the edge of instability \cite{YuHu}. To facilitate the comparison among regions, for each of them the curve has been normalized by the maximum eigenvalue of the covariance matrix.}
\new{ (\textbf{E}) Long time-window covariance eigenvalues distribution for an example region (MOp, blue histogram), together with the best-fitting Marchenko-Pastur distribution (green line, fitted parameter $\hat{\alpha}=1.00$) and the best-fitting covariance eigenvalues distribution for a linear-rate model of randomly connected neurons (red line, fitted parameter $\hat{g}=0.95$). (\textbf{F}) Fitted values of $g$ and $\lambda_{max}$ (including errorbars, calculated as in D) for all the considered regions. Inset: deviation of the empirical distribution to the Marchenko-Pastur (MP) distribution versus deviation to the theoretical distribution for a recurrent network (RC) of linear rate neurons close to the edge of instability.}
\label{Distance_Criticality}}
\end{figure*}

In particular, to compute meaningful covariances one assumes that neural activity is stationary, a condition that typically holds during recordings of \newd{background,} resting-state type of activity, but not as much in task-related activity, for which network activity is inherently input-driven and non-stationary. Therefore, we restrict our forthcoming analyses to resting-state data. \new{We also notice that the values of spike-count covariances depend on the window sizes over which such counts are measured \cite{Helias}. In the forthcoming results, we took sampling windows of $T=1s$, which are sufficient for autocorrelations to decay while maximizing the number of samples over which spike-count covariances are computed
(see Fig.\ref{Distance_Criticality}A an Table S1 in the SI). Using such pairwise spike-count covariances measured
in a given region, we  consider two alternative methods to infer the distance of its dynamical state to the edge of instability:}

\emph{(i) Estimations from the statistics of spike-count-covariance distributions.}   Calling  the  mean value of the distribution of such spike-count covariances  $\bar{c}$ and its standard deviation $\delta c$, the largest eigenvalue of the connectivity matrix, $\lambda_{max}$ ---whose closeness to $1$ marks the distance to the edge of instability, \new{thus gauging} the dynamical regime--- can be determined from the following expression \cite{Helias}:
\begin{equation}
    \lambda_{max} = \sqrt{1-\sqrt{\dfrac{1}{1+N\Delta^{2}}}},
    \label{lambda_max}
\end{equation}
where  $N$ is the total number of neurons in the population, $\Delta = (\delta c)/\bar{c}$. 
\new{We remark that the derivation of Eq.(\ref{lambda_max}) is based on linear response theory and relies on tools from spin-glass theory applied to  networks of spiking neurons, which assumes the stationarity of the spiking statistics \cite{Helias}. An Augmented Dickey-Fuller test was performed to verify in each case the validity of this assumption (see Table S1 and section \textit{Stationarity of spiking statistics} in Extended Methods of the SI)}.

As illustrated in Fig.\ref{Distance_Criticality}B (for one of the selected regions), the \newd{resulting} distributions \new{of pairwise spike-count covariances} turn out to be sharply-peaked around $0$, but they exhibit significant tails, revealing the presence of heterogeneously correlated pairs in all areas. 

\new{From these spike-count covariances, one could then easily estimate the distance to the critical point for the neural dynamics in each region (see Table S1 and Fig.\ref{Distance_Criticality}F). Let us remark, however, as it follows from Eq.\ref{lambda_max}, that the value of $\lambda_{max}$ strongly depends on the number of neurons being recorded and, since the available empirical data heavily sub-samples each region, the aforementioned approach would actually underestimate the real value of $\lambda_{max}$, which becomes  much closer to $1$ in the limit of tens of thousands of neurons. In Fig.\ref{Distance_Criticality}C we plot the values of $\lambda_{max}$ as a function of the number of recorded cells, showing that the larger the neural population size, the larger the value $\lambda_{max}$.

Moreover, because of this dependence with the system size, it becomes difficult to compare the estimated values across regions with different numbers of neurons recorded. To avoid this limitation, we computed for each experiment and region the normalized covariance width, $\Delta t$, and then applied Eq.\ref{lambda_max}, extrapolating to a common number of neurons $N=10^4$. In Fig.\ref{Distance_Criticality}D we show the distance to the critical point thus estimated, with errorbars computed as the standard deviations across experiments in each region. Notice that most values lie on a very narrow window between $0.96$ and $0.99$, with a mean value $\overline{\lambda}_{max} = 0.978 \pm 0.009$, close to the edge of instability.}

\newd{Direct application of Eq.\ref{lambda_max} for the $16$ considered regions in the mouse brain leads to the \new{estimations} for the maximum eigenvalue, $\lambda_{max}$, in each case, as summarized in Fig.\ref{Distance_Criticality}D (the corresponding average values and errors across experiments are collected in Table S1 of the SI). Notice that most of them lie on a relatively narrow window between \new{$0.81$ and $0.93$, with a mean value $\overline{\lambda}_{max} = 0.84 \pm 0.08$,} close to the edge of instability.

In any case, let us stress that the real dynamical state of all regions is actually closer to criticality than what is suggested by Fig.\ref{Distance_Criticality}D: indeed, it follows from Eq.\ref{lambda_max} that the value of $\lambda_{max}$ strongly depends on the number of neurons in the region and, since the available empirical data heavily sub-samples each region, the previous approach actually underestimates the real value of $\lambda_{max}$, which becomes  much closer to $1$ in the limit of tens of thousands of neurons. \new{In Fig.\ref{Distance_Criticality}C we plot the values of $\lambda_{max}$ as a function of the number of recorded cells and show that the larger the neural population size, the larger the value $\lambda_{max}$.} Thus\newd{, as shown in Fig.\ref{Distance_Criticality}C,} all the analyzed regions become much closer to the edge of instability once the previous approach is extrapolated to sufficiently large, more realistic, neural population sizes (of the order of $10^4$ neurons).}

\new{
\emph{(ii) Estimations from the eigenvalue spectrum.} 
Hu and Sompolinsky \cite{YuHu} have recently derived an analytical expression for the distribution of eigenvalues of the spike-count covariance matrix for a recurrent random network of linear rate neurons, whose dynamics follows the equation:
\begin{equation}
     \dot{x_i}(t)=-x_i(t)+ g \sum_{i=1}^N J_{ij} x_j +\xi_i(t)
    \label{rate}
\end{equation}
with $\langle\xi_i(t)\xi_j(t+\tau_0)\rangle=\sigma^2\delta_{ij}\delta(\tau_0)$ and $J_{ij}\sim \mathcal{N}(0,1/N)$. In the model, the parameter $g$ sets the overall connection strength, with $1-g$ marking the distance to the edge of instability \cite{YuHu}. This result allows one to estimate the network dynamical state by fitting the actual empirically-determined spectrum to the theoretical distribution as a function of  $g$ (see Extended Methods in the SI and \cite{YuHu}). In particular, Fig.\ref{Distance_Criticality}E shows the best fit ($g=0.95$) of the empirically-determined eigenvalue distribution for the MOp region to the Hu-Sompolinsky distribution, together with a  (much worse) fit to the Marchenko-Pastur distribution (expected for random uncorrelated time series), thus underlying the non-trivial structure of the observed spike-count covariances, which stem from recurrent quasi-critical interactions. A summary of the inferred $g$-values for all areas is shown in Fig.\ref{Distance_Criticality}F (see also Table S1), which further illustrates the strong similarity between the results obtained with the two employed methods, when the original number of neurons in each experiment is considered.} 

\newd{Finally, let us mention that a possible connection between the measured distance to criticality and the functional role of cortical areas is suggested in the Discussion Section.}

\new{
\subsection*{Non-trivial scaling features emerge in the RG analysis of recurrent random-network models at the edge of instability}

To close the loop, we wondered whether a simple model of randomly-coupled linear units, as the one defined by Eq.(\ref{rate}), is able to reproduce also other non-trivial scaling features as revealed by phenomenological-RG analyses. In Fig.S10 of the SI we show evidence that such recurrent neural networks driven by noise   generate patterns of activity  that 
reproduce many of the non-trivial scaling features emerging out of the RG analyses when they are posed close to the edge of instability (e.g., $g = 0.95$), while such non-trivial features are lost in the subcritical regime (where trivial Gaussian scaling emerges). In particular, the  values of the exponents at the edge of instability ($\alpha \approx 1.28$, $z \approx 0.21$ and $\mu \approx 0.78$) are remarkably similar to the ones measured across regions in the mouse brain (see Table S2). The observation is rather striking given that these exponent values come from a linear model of randomly connected neurons, while real data are, most likely, generated by a more complex non-linear dynamics on heterogeneous networks.
A recent work on universal aspects of brain dynamics \cite{Tiberi} might help shedding light onto this result.
}

\subsection*{Non-universal, input-dependent exponents in task-related activity}

Saying that external stimuli shape neural correlations in information processing areas is, to a certain extent, an obvious statement. However, the fact that the spectrum of the activity covariance matrix follows a very simple mathematical rule that ultimately ensures the smoothness of the internal representation of inputs ---as mathematically proved by Stringer \emph{et al.} \cite{Stringer-Nature}--- is quite remarkable. 
In particular, these authors showed that the neural representation for a $d$-dimensional (visual) input ---that is to be encoded collectively in the neural activity of the primary visual cortex (VISp)--- is constrained by the requirement of  smoothness of the representation (i.e., continuity and differentiability of the associated representation manifold)  \cite{Stringer-Nature}.
Being more specific, for these conditions to hold, the spectrum of the covariance matrix needs to decay as a power-law with an exponent $\mu$ that must be greater than $1$ for continuity, and greater than $1 + 2/d$  ---where $d$ is the dimension of the input ensemble (see \textit{Datasets} section in Extended Methods of the SI)--- for differentiability \cite{Stringer-Nature}. Thus, for sufficiently "complex" inputs, i.e. with large embedding dimensionality ($d$), this exponent is constrained to take a value arbitrarily close to (but larger than) unity, $\mu \gtrsim 1$.

On the other hand, in the previous RG analyses, we found values of $\mu$ consistently smaller than $1$ ---both under resting conditions and in task-related activity--- for all the considered areas (see Fig.\ref{Eigenvalues_Spont}), \new{in agreement with \cite{Bialek-PRL}, but} in seeming contradiction with the predictions of Stringer \emph{et al.} \cite{Stringer-Nature}. How come that we report values $\mu<1$ in all areas including sensory information encoding ones, even when the mouse is exposed to external stimuli? Does it mean that stimuli encoding violates the requirements for efficient representation put forward by Stringer \emph{et al.} \cite{Stringer-Nature}?

At the core of this seeming paradox lies a data-processing method proposed by Stringer {\emph et al.} that allows one to extract the input-only related covariances from the overall "raw" covariance matrix \cite{Stringer-Nature}. The approach stems from the idea that population activity can be decomposed into an  input-related (or "input-encoding") subspace, which spans input-only related activity and a complementary space ---orthogonal to the former one--- which captures  the remaining activity \cite{Rule}. In a nutshell, the so-called cross-validated PCA (cv-PCA) method consists in repeating twice the very same experiment and comparing the two resulting raw covariance matrices; this comparison allows one to infer which part of the covariance is shared by the two matrices and is, hence, input-related (see secion 6 in the SI) and which complementary part stems from unrelated background activity \cite{Stringer-Nature}.
\new{Unfortunately, given the nature of such an experimental protocol, it is not possible to apply cv-PCA to the dataset of Steinmetz {\emph et al.} \cite{Steinmetz} extensively employed above.}

\new{However, to} \newd{To} prove that our RG results above (consistent with  $\mu <1$) are not in \newd{blatant} contradiction with the ones by Stringer {\emph et al.}   ($\mu \geq 1$), \new{we have extended} the cv-PCA method (as explained in detail in section 6 of the SI) to be able to actually extract from  empirical data in \cite{Stringer-Nature} not just the input-related covariance matrix but also the time-series of input-related neural activity. For this, the overall activity $x(t)$ of a given neuron at time  $t$ is projected into two separate sub-spaces, i.e. decomposed as:
\begin{equation}
    x (t) = \psi (t) + \epsilon(t)
    \label{decomposition}
\end{equation}
where $\psi(t)$ describes its input-related activity and $\epsilon(t)$ stands for the remaining ``orthogonal''  activity, responsible for trial-to-trial variability. This decomposition allows us to perform separate RG analyses to input-related and background activity data. As a  consistency check, we also verified that the covariance matrix eigenspectrum associated with $\psi(t)$ has no significant difference with the one obtained from the standard application of cv-PCA analyses in \cite{Stringer-Nature} (see Fig.S15 in the SI for further details).


\begin{figure}
\begin{center}
\includegraphics[width=11cm]{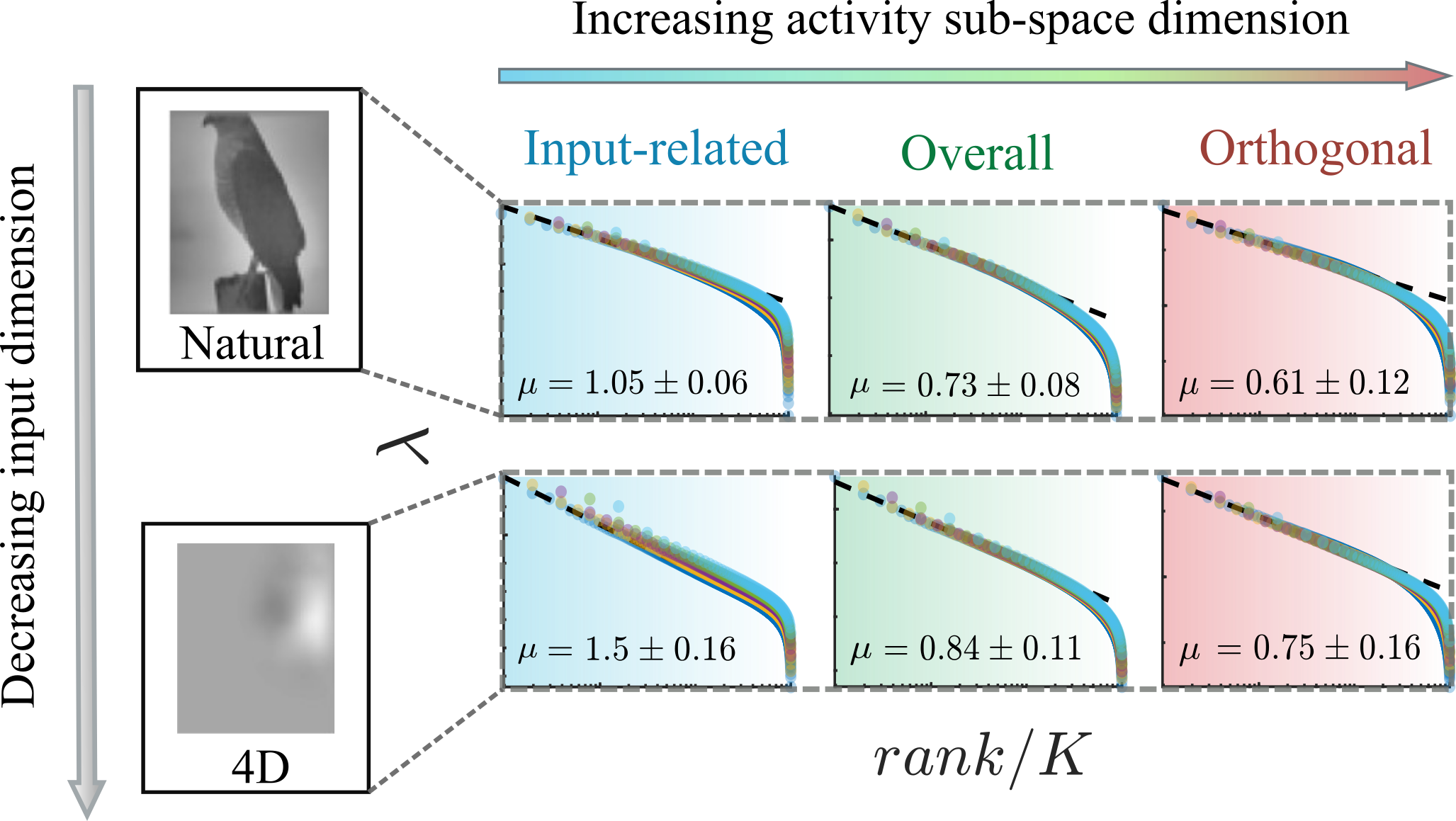}
\caption{Diagram showing the two observed trends in the power-law exponent $\mu$, which characterizes the decay of the covariance-matrix eigenvalues as a function of their rank. $\mu$ decreases with the complexity of the input when activity is projected into the task-encoding  subspace (also called representation manifold). On the other hand, for a fixed type of input, the exponent decreases with the proportion of background, "noisy" activity in the data, which lies in a higher-dimensional subspace orthogonal to the representational manifold. Natural and low-dimensional images examples have been adapted from \cite{Stringer-Nature}.
\label{Trends}}
\end{center}
\end{figure} 

We now proceed to show the main results of applying the phenomenological RG approach over
\new{the data of \cite{Stringer-Nature} considering:} \emph{(i)} the raw \new{timeseries} $x(t)$, \emph{(ii)}  the input-related activity $\psi(t)$, and \emph{(iii)}  the background activity $\epsilon(t)$, with the results averaged over three different mice.  

\emph{(i)} Analyzing the raw data (i.e. $x$\newd{'s} variables) one observes again exponent values ($\alpha = 1.49 \pm 0.08$ and $\mu = 0.73 \pm 0.08$) in agreement with the previously-reported quasi-universal values for different mouse-brain areas (see Fig.\ref{Trends}, together with Fig.S7 and Table S1 in the SI), \new{highlighting the robustness of our results for very different recording techniques.}

\emph{(ii)} Considering the input-evoked activity ($\psi$\newd{'s}) one finds that the exponents $\alpha$ ---which was rather robust when measured from the raw-activity data--- is significantly altered (see SI, Fig.S7). More importantly, a significant increase in the exponent $\mu$ is observed with respect to the raw-data analyses; it now respects the theoretical boundary $\mu > 1 + 2/d$ (with $d \longrightarrow \infty$ in natural images) for the smoothness of the representation manifold (see Fig.\ref{Trends}). In addition, the value of  $\mu$ obtained from our RG analyses decreases as the input dimensionality grows, in agreement with the theoretical results and empirical findings in \cite{Stringer-Nature}. 

\emph{(iii)} Finally, considering only the residual "orthogonal" parts ($\epsilon$\newd{(t)'s}) we found slightly smaller values of $\mu$ with respect to the raw data analyses for the two types of inputs considered  (see Fig.\ref{Trends}). We notice, however, that this background activity is not to be confused with simple "white noise" as it exhibits correlations with a non-trivial power-law spectrum. When looking at the exponent $\alpha$ for the variance inside block-neurons, we did not observe a significant change with respect to the overall-activity case (Fig.S7 in SI). This further supports the idea that the observed scaling exponents in the overall raw data are dominated by the higher-dimensional, background activity.

To summarize the previous results,  Fig.\ref{Trends} recaps the two observed trends in the scaling exponent $\mu$. On the one hand ---for data obtained using images of different intrinsic dimensionalities (see Extended Methods in SI)---  we found the relationship predicted and  observed by Stringer \emph{et al.} in \cite{Stringer-Nature}, namely, that for the input-related activity $\psi$\newd{(t)'s}, the value of $\mu$ decreases with the complexity of the input signal, with a lower limit of $\mu=1$ for high-dimensional  inputs. On the other hand, the exponent $\mu$ decreases with the relative weight of background activity in the data: the more input-related activity that is projected away, the \newd{more "noisy" (}flatter{\newd)} the spectrum of the remaining orthogonal space.

As a side note, let us mention that  ---owing to the nature of the experimental setup \new{in \cite{Stringer-Nature}}, in which stimuli are interspersed with grey-screen inter-stimulus intervals --- computation of actual dynamical correlations are not meaningful in this case, so that we  purposely left aside the study of the dynamical scaling exponent $z$. Likewise, the transformations carried on the overall data to extract the input-related and orthogonal activity do not necessarily preserve the biological significance of the zero-activity, making the computation of the "free-energy" exponent $\beta$ pointless.

Thus, in summary, application of the phenomenological RG procedure to the data for \newd{the} VISp \newd{visual cortex} \new{of Stringer {\emph et al.} \cite{Stringer-Nature}} reveals that the overall activity exhibits clear signatures of scale invariance, sharing its quasi-universality with the previously analyzed regions, which relied on a different dataset. Nevertheless, this overall activity can be decomposed into input-related and noisy/orthogonal activities: the scaling exponent for the covariance matrix in the first case obeys the mathematical constraints derived by Stringer {\emph et al.}, thus solving the seeming contradiction with the spectrum of the overall data.

\section*{Discussion}

Understanding how the brain copes with  inputs from changing external environments, and how  information from such inputs is transmitted, integrated, processed, and stored in a physical substrate consisting of noisy neurons ---exposed also to a stream of other overlapping inputs--- is one of the major challenges in neuroscience. The fast development of revolutionary experimental and observational techniques  as well as of novel theoretical insights have resulted in huge advances in this field in recent years. 

On the one hand,  novel technologies allowing for  the simultaneous recording of thousands of neurons, pave the way to quantitative analyses of brain activity with unprecedented levels of resolution and detail, making it possible, for the first time, to discriminate between different overarching theories.

Here, we have taken advantage of high-throughput data, together with state-of-the-art theoretical approaches  to analyze neuronal activity across regions in the mouse brain. One of our chief objectives was to make progress in elucidating whether the so-called "criticality hypothesis" ---in some of its possible formulations--- is supported by empirical data.  \newd{and, thus, can be further pursued as an overarching principle to rationalize how the intrinsic  dynamics of neural networks in the brain sustains information processing.} This \newd{ambitious} goal has been tackled in two steps. 

The first step was to assess the presence or absence of scale invariance or "scaling" in neural data, for which we extensively rely on the phenomenological RG approach recently proposed by Meshulam {\emph et al.} \cite{Bialek-PRL,Bialek-largo,Bradde}: our analyses here confirm the existence of strong signatures of scaling in all the analyzed brain regions, with exponent values taking quasi-universal values with only relatively small variations across areas \cite{Steinmetz}. The level of universality is not as precise as in critical phenomena  i.e. in "non-living matter" \new{(though, even in Physics, critical exponents can take non-universal, continuously-varying values depending, for instance, on structural heterogeneity/disorder (see e.g. \cite{Moretti}).} 
Further work is needed to better understand the origin and functional meaning  of this variability. 

The second step was to scrutinize whether the empirically observed scale invariance stems from criticality or not. Thus, the next sections are devoted to the discussion of this hypothesis at the light of our presented results.

\subsection*{Criticality versus latent dynamical variables}
Importantly, as already mentioned, recent works \cite{Nicoletti,Nemenman-PRL} have shed doubts on the possible relation between the  empirically-found scaling in neural recordings and actual criticality, as originally suggested by Meshulam {\emph et al.} \cite{Bialek-PRL,Bialek-largo}. In particular, Morrell {\emph et al.} constructed a very simple model of binary neurons which, \new{being uncoupled,} cannot possibly exhibit collective behavior such as phase transitions or criticality \cite{Nemenman-PRL}. In their setup, all individual neurons are exposed to a large set of shared external, time-correlated inputs, so-called "latent dynamical variables" or simply "hidden variables". Surprisingly enough, such a toy model is able to reproduce ---for a relatively broad region of the parameter space--- the RG scaling with non-Gaussian activity distributions as well as a set of  exponents roughly compatible with \newd{that} \new{those} in actual neural networks \cite{Bialek-PRL,Bialek-largo}. This suggests that the empirically observed scale invariance in neural recordings could possibly \new{have little to do} with intrinsic reverberating critical dynamics, but it can emerge as an evoked response to shared external drivings.

Let us recall that this type of dichotomy for the interpretation of scaling ---between critical behavior on the one hand and a superposition of the simpler effects stemming from hidden variables on the other---  is a common theme in different branches of science, being for example at the roots of discussions about the meaning of the Zipf's law in neural data \cite{Newman,Latham,Mehta,Marsili-2019}.

\newd{Furthermore, as already emphasized by Hu and Sompolinsky, a long tail in the distribution of the eigenvalues of the spike-count covariance matrix ---which is one of the chief properties we find in the analyzed data--- is a distinct feature of correlations arising from nearly-unstable recurrent network dynamics \cite{YuHu}: it cannot be possibly reproduced by a network of uncoupled neurons and, thus, it cannot be merely an effect stemming from superposed hidden inputs}

\new{Here,  by employing a variety of tools, we have concluded that all the empirically analyzed brain regions lie ---to a greater or lesser extent--- at the edge of instability, i.e. in the vicinity of a critical point separating stable from linearly unstable phases. Moreover, we have also shown that a random network of linear units tuned close to the edge of instability suffices to reproduce non-trivial scaling features, with remarkably similar exponent values without the explicit need of external  fields.}

Let us caution, however, that the previous results do not imply that external (latent) dynamical inputs may not have an impact on the \new{recurrent dynamics nor on the observed scaling exponents. External inputs contribute to set the network working point at which covariances (as well as the stability matrix of the linear-rate model) are computed. Furthermore, the observed  differences in exponent values across regions could stem from diverse exposures to latent fields from other areas.} 

Further empirical and theoretical studies would be required to advance in this direction; in particular, developing more elaborated models including explicitly both, near-critical recurrent dynamics and shared latent variables, as well as analyzing how the relative weights of both contributions shape scaling exponents remain open goals for future research.

\newc{Finally, a particularly relevant open question is whether the observed differences 
across regions can be related to the specific functional role of each area. Even if we do not have a clean-cut answer to this, let us make the following observation. One parsimonious measure of the role of a (mouse) brain area is its hierarchical score as recently determined \cite{JulieHarris}, with low (high) scores corresponding to sensory (higher-level) areas.  Results in Fig.\ref{Distance_Criticality}D allow us to observe  that more sensory areas such as primary cortices MOp and SSp (with low scores; see e.g. Fig.6 in \cite{JulieHarris}) tend to operate closer to the edge of instability than secondary cortices (such as MOs) or areas in the prefrontal cortex (such as ORB, PL and ACA). This seems to suggest that there could exist a relationship between the dynamical regime of a given area and itsu hierarchical score, with low-score regions being more "critical".
On the other hand, we do not observe a clear and consistent trend in thalamic regions such as LGd, LP, MG and PO.
Thus, more thorough and comprehensive studies would be needed to draw solid conclusions on this inspiring possible connections.}

\subsection*{Edge of instability and optimal representations}

By applying the phenomenological RG procedure to the data for VISp from Stringer {\emph et al}. \cite{Stringer-Nature}, we found that the overall activity exhibits clear signatures of scale invariance and shares its quasi-universality with the previously analyzed regions relying on a different dataset, solving a seeming contradiction between our own results for quasi-universal scaling across brain regions and those in \cite{Stringer-Nature}. However, the projection of the overall or "raw" activity into input-representing and complementary/orthogonal-space activities allowed us to conclude that the 
scaling exponent determined from RG analyses in the first case obeys the mathematical constraints derived by Stringer {\emph et al}. \new{Understanding how the brain performs this task, i.e. how it separates signal from noise is an open fundamental problem (see e.g. \cite{Semedo}).}

Finally, let us emphasize that, given that all the analyzed  regions are near critical, they are bound to exhibit power-law decaying covariance-matrix spectra. Thus, we conjecture that critical behavior creates the broad range of covariance scales, needed for neural networks to support optimal input representations with power-law decaying eigenvalues. More research is needed to confirm this conjecture and put it on firmer ground. 

\new{To further  illustrate the possible relationship
between power-laws in the spectra of covariance matrices and optimal input representations, let us mention that, in a related work, we have recently designed and analyzed a simple machine-learning model based on the paradigm of \emph{reservoir computing} \cite{Morales}} to analyze this problem in a well-controlled example. \new{This model consists of a recurrent network of coupled  units/neurons, which receive shared external inputs, giving rise to reverberating activity within the network or "reservoir".} Contrarily to other machine-learning paradigms, the internal synaptic weights remain fixed during the training process: only a smaller subset of links connecting to a set of readout nodes change during training, which makes reservoir-computing a versatile tool for diverse computational tasks \cite{reservoir}. Inspired by the experiments of Stringer {\emph et al.}, we trained the network on an image classification task. We refer the interested reader to \cite{Morales} for further details. For our purposes here, it suffices to recall that the best performance is obtained when the tunable control parameters are set in such a way that the overall dynamical state is very close to, but below, the edge of instability. Moreover, within such an operational regime, the spectrum of the covariance matrix obeys the mathematical requirement for optimal representations, i.e., $\mu \gtrsim 1$, as observed for actual neural networks \cite{Stringer-Nature}.

We find quite suggestive that such a relatively-simple artificial neural network becomes optimal in a regime that shares crucial statistical properties of the covariances with actual neural networks in the (mouse) brain. Thus, we believe that this machine-learning model may constitute a well-controlled starting point to further investigate the interplay between internal dynamics and external shared inputs in more realistic models of brain activity and to scrutinize optimal input representations. In any case, this seems to confirm that scale-invariant covariance spectra, as a naturally emerging property of critical systems, constitute an excellent breeding ground for optimal information storage.

\subsection*{Avalanche criticality versus edge-of-instability} 

As already discussed, we have found strong empirical evidence of critical behavior in the sense of vicinity to the edge of instability across brain regions. This type of behavior ---called traditionally "edge of chaos" or "type-II criticality" in \cite{Helias}--- has long been (since the pioneering works of Langton and others \cite{Langton,BerNat}) theoretically conjectured to be crucial for information processing in natural and artificial neural  networks. In this case, edge-of-instability systems  are characterized by the presence of many modes that are close to become unstable, and thus there is a large repertoire of possible dynamical states that can be excited, opening many possible channels to information processing and transmission in real time \cite{BerNat,Boedecker,RMP}.

However, in the analyzed neural recordings there are not large fluctuations in the overall level of global activity across time, in agreement with what observed for the motor cortex of awake macaque monkeys in \cite{Helias} and with the expectation for  "asynchronous states" in balanced networks \cite{Renart}.  Nevertheless, it is \new{noteworthy}, that in some other empirical observations diverse levels of temporal variability in collective firing rates \new{(e.g. gamma oscillations)} have been also reported together with the possibility of bursts and avalanching behavior \cite{BP2003,Petermann,Fontenele}. Actually, as  stated in the Introduction, much attention has been paid to "avalanche criticality" (referred as "type-I criticality" in \cite{Helias}), which is associated with networks in which there is an overall-activity mode about to become unstable, thus generating scale-free avalanches of activity, while the rest of modes are stable. Importantly, scale-free avalanches can also occur at the edge of synchronization phase transitions
where the dominant mode becomes oscillatory
\cite{LG,Zhou-Hopf,Buendia}.

It should be emphasized that both types of criticality are not mutually exclusive as, in principle, it is possible to have a whole set of eigenvalues, including the overall-activity one, at the edge of instability. Similarly, the \new{network} could be at the edge of becoming collectively oscillatory (as recently proposed for large-scale brain networks in e.g. \cite{LG,Zhou-Hopf}) if the leading eigenvalue crosses the instability threshold with a non-vanishing imaginary part and at the same time have many other modes near the edge of instability as found here. As an illustrative example, let us mention that a computational model of spiking neurons, including both "asynchronous states" and a synchronization phase transition characterized by scale-free avalanches, has been recently proposed and analyzed \cite{Zhou-Hopf}.
Thus, it seems a priori \newd{perfectly} possible to construct computational models exhibiting both types of criticality in which the system can shift between different regimes depending on its needs. For instance, \new{asynchronous states near the edge-of-instability} support local computational tasks, while emerging synchronous oscillations are useful for information transfer and reliable communication with distant areas \cite{Bassett-oscillations}. In our view, it is likely that actual brain networks  exploit these two complementary ways of being critical to achieve diverse functional advantages for different tasks, possibly by shifting their dynamical regimes in response to stimuli.

\vspace{1cm}

\section*{Conclusions} 
We have  developed a \newd{unified theoretical} \new{synergistic} framework which relies on recently proposed breakingthrough approaches, but that also extends and combines them \newd{in a synergistic way, allowing us} to analyze state-of-the-art recordings of the activity of many neurons across brain regions in the mouse. We find that all regions exhibit scale invariance and that all of them operate, to a greater or lesser extent, in a critical regime at the edge of instability.
Moreover, we have shown that the resulting scaling in the spectrum of covariances can have very important functional applications for information storage, as it facilitates the generation of optimal input representations.
It is our hope that the present work stimulates further research on the remaining open questions and helps advance towards a more comprehensive understanding of the overall dynamics of brain networks and their emerging computational properties, as well as to disentangle universal and non-universal aspects across regions and behavioral states.



\clearpage
\newpage
\section*{Appendix}
\subsection*{Phenomenological renormalization-group (RG) approach}

Let us  briefly outline the phenomenological RG approach introduced in \cite{Bialek-PRL} and \cite{Bialek-largo}. Given a set of $N$ neurons, the empirically-determined activity of the $i$-th neuron at a given time $t_j$ is denoted by $\sigma_{i}(t_j)$, where $j\in(1,T)$ labels a discrete number of non-overlapping time bins. Determining the most meaningful size of such time bins is an important technical aspect (see next section); here, we
just assume that such an optimal time discretization is given.

As discussed in the main text, a criterion of maximal pairwise correlation is employed to group neurons together at each step $k$ of the RG procedure. In particular, one considers the Pearson's correlation coefficients
\begin{equation}
C_{ij}^{(k)}= 
\langle \delta x_{i}^{(k)}\delta x_{j}^{(k)}\rangle /
\sqrt{
     \langle       (\delta x_{i}^{(k)}  )^{2} \rangle 
\langle (\delta x_{j}^{(k)})^{2}\rangle},
\end{equation}
where $\delta x_{i}^{(k)}=x_{i}^{(k)}-\langle x_{i}^{(k)}\rangle$
and $x_{i}^{(k)}$ is the activity of the block-neuron $i$ at step $k$ of the coarse-graining (we identify $x_{i}^{(0)}\equiv\sigma_{i}$ as the activity of neuron $i$ before coarse-graining), while averages are computed across the available discrete time steps.
At the beginning of each RG step, the two most correlated neurons, $i$, and $j_{*i}$, are selected and their activities are added into a new coarse-grained variable:
\begin{equation}
x_{i}^{(k+1)}=z_{i}^{(k)}\left(x_{i}^{(k)}+x_{j*i}^{(k)}\right).
\end{equation}
where the normalization factor $z_{i}^{k}$ is chosen in such a way that the average non-zero activity of the new variables $x_{i}^{k+1}$ is equal to one. Notice that, owing to such a normalization criterion, activity values are not constrained to fulfill $x_{i}^{k}(t)<1$. Then one proceeds with the second most-correlated pair of neurons and so on, until a set of $N_{k}$ coarse-grained "block neurons", each containing the summed activity of $K=2^{k}$ original neurons, has been constructed. 

Iterating this procedure, after $k$ steps there remain only $N_{k}=N/2^{k}$ coarse-grained variables or "\emph{block-neurons}", $\{x_i^k\}_{i=1,2...N_k}$, each recapitulating the activity of $K=2^{k}$  individual neurons. To figure out whether a fixed point of the RG flow exists, one can study the evolution of the probability density function for the activity of the coarse-grained variables, $P_{K}(x)$. Following \cite{Bialek-PRL}, we separated $P_{K}(x)$ for block-neurons of size $K$ in two components:  the probability of being silent, $S_{K}$, and the probability $Q_{K}(x)$ of having non-zero activity $x$:
$P_{K}\left(x\right)=S_{K}\delta\left(x\right) + (1-S_{K})Q_{K}\left(x\right)$.
Trivially, if the original neurons were statistically independent, one would expect (as a direct consequence of the central limit theorem) to drive the activity distribution $Q_{K}$  towards a Gaussian fixed-point of the RG flow. 
As pointed out in \cite{Bialek-PRL}, the RG convergence to a non-Gaussian fixed-point (i.e. invariance of the distribution across RG steps) reveals a non-trivial structure in the data.

Another quantity of interest is the variance of the activity distributions as a function of the size of the block neurons $K$:
\begin{equation}
M_{2}(K)=\dfrac{1}{N_{k}}\sum_{i=1}^{N_{k}}\left[\left\langle \left(\sigma_{i}^{(k)}\right)^{2}\right\rangle -\left\langle \left(\sigma_{i}^{(k)}\right)\right\rangle ^{2}\right]
\end{equation}
where $\sigma_{i}^{(k)}$ is the summed activity of the original variables inside the cluster. Notice that, for totally independent variables, one would expect the variance to grow linearly in $K$ (i.e., $M_{2}(K)\propto K$), whereas if variables were perfectly correlated $M_{2}(K)\propto K^{2}$. Non-trivial scaling  is therefore characterized by
$M_{2}(K)\propto K^{\alpha}$
with a certain intermediate value of the exponent $1<\alpha<2$. 

On the other hand, 
$F_{k}=-\log\left(S_{K}\right)$
defines a sort of "free-energy" for the coarse-grained variables at the $k$-th RG step \cite{Bialek-PRL}. As  more and more of the initial variables $\sigma_{i}$  are grouped into cluster variables $x_{i}^{(k)}$, one would expect that the probability of having "silent" block-neurons (i.e., the probability that all neurons inside a cluster are silent) decreases exponentially with the size $K$ of the clusters, leading to:
$F(K)\propto K^{\beta}$
where $\beta=1$ for initially independent variables. 
 
One can also wonder whether there is some type of self-similarity in the dynamics at coarse-grained scales. Given that, commonly, fluctuations on larger spatial scales relax with a slower characteristic time scale, we should expect the time-lagged Pearson's correlation function (or simply autocorrelation function) of the coarse-grained variables to decay more slowly as we average over more neurons. In particular, for step $k$ of the RG flow, one has:
\begin{equation}
C^{(k)}(t)=\dfrac{1}{N_{k}}\sum_{i=1}^{N_{k}}\frac{\langle  x_{i}^{(k)}(t_{0})x_{i}^{(k)}(t_{0}+t)\rangle - \langle x_{i}^{(k)}\rangle^{2}}
{\langle {(x_{i}^{(k)})}^{2}\rangle - \langle x_{i}^{(k)}\rangle^{2}}
\end{equation}
Assuming that correlations decay exponentially in time with a characteristic time scale $\tau_{c}^{(k)}$  (i.e., $C^{(k)}(t)=e^{-t/\tau_{c}^{(k)}}$) at each coarse-graining level, dynamical scaling implies that the average correlation function collapses into a single curve when time is re-scaled by the characteristic time scale:
$C^{(k)}(t) = C(t/\tau_{c}^{(k)})$
and that this time scale obeys scaling with the cluster size:
$\tau_{c}(K) \propto K^{z}$
where $z$ is the dynamical scaling exponent. Finally, as argued in \cite{Bialek-PRL}, if correlations are self-similar then we should see this by looking inside the clusters of size $K$. In particular, the eigenvalues of the covariance matrix (i.e., the propagator, which is scale-invariant at the fixed-point of the RG in systems with translational invariance \cite{Bialek-PRL}) must obey a power-law dependence on the fractional rank:
$\lambda = B\left(K/rank\right)^{\mu}$,
where $B$ is a constant and $\mu$ a decay exponent.

\vspace{0.5cm}
\new{ We estimate the goodness of each power-law fit by calculating the R-squared value, comparing too with an equivalent exponential fit (see Fig.S9 in the SI) \cite{Clauset}. For the probability density of eigenvalues we compute the log-likelihood ratios between the estimated power-law and alternative exponential and lognormal distributions (see Extended Methods and Tables S1 and S2 in the SI). }

\newd{To compute an error on the estimate exponents we follow the approach in \cite{Bialek-largo}: for each region, we reproduce the RG analysis over random quarters (batches) of the data, with points remaining in order within these intervals to respect temporal correlations. We then estimate for each batch the scaling exponents as the slope of the best fit in a log-log scale \cite{Clauset}. Our final estimation and error for any scaling exponent within a experiment is then computed as the mean and standard deviation across all batches, respectively} \new{

Each exponent is expressed as
  $e=\bar e + MAE +\sigma$,
where $\bar e$ is the average across different experiments (possibly from different mice), $MAE$ is the mean-absolute-error, computed as the average across experiments of the experiment-specific errors measured over split-quarters of data, and $\sigma$ is the standard deviation across experiments (see Tables S1 and S2 in SI).}


\new{
\subsection*{Measuring covariances}
Here, we use a general definition of covariance, as described by the following equation:
\begin{equation}
c_{ij}^{s} = \left\langle 
(x_{i}-\left\langle x_{i}\right\rangle)
(x_{j}-\left\langle x_{j}\right\rangle
\right\rangle.
\label{signal}
\end{equation} 
During the RG analysis and in estimations of the distance to the edge of instability, the variable $x_{i}$ represents the number of spikes in a time bin of width $\Delta t$, with averages taken over all timebins, so that $c_{ij}$ measures the pairwise correlation of spike-count responses to repeated presentations of the same stimulus (or, in our case, repeated sampling of resting-state activity under identical behavioral conditions). Throughout this article, we referred simply as "covariance" when correlations were computed in the RG analysis using the time bin given by the geometric mean of neurons' ISIs, which renders time series where the average non-zero bin is populated by only one spike. In contrast, the term "long-time-window" or "spike-count" covariance is left for the distance to criticality analysis, in which $\Delta t=1$s and the average non-zero bin in the time-series contains between $3$ and $7$ spikes, depending on the region. We notice that the later can be written as the time integral of the time-lagged covariance $c_{ij}(\tau)$ \cite{Helias}:
\begin{equation}
    c_{ij}^{n} = \lim_{\Delta t_{0}\longrightarrow \infty} \int_{-\Delta t_{0}}^{\Delta t_{0}} \dfrac{\Delta t_{0}-\tau}{\Delta t_{0}}c_{ij}^s(\tau)\, d\tau.
\end{equation}
which, loosely speaking, removes time-dependent effects and puts the emphasis onto  pairwise heterogeneities. 
}

\vspace{0.5cm}
\new{
\subsection*{Extended Methods} See the "Extended Methods" section in the Supplementary Information text for additional details.}

\vspace{6pt} 

\subsection*{Acknowledgements}
We acknowledge the Spanish Ministry and Agencia Estatal de investigaci{\'o}n (AEI) through 
Project of I+D+i Ref. PID2020-113681GB-I00, financed by 
MICIN/AEI/10.13039/501100011033 and FEDER “A way to make Europe”, as well as the Consejer{\'\i}a de Conocimiento, Investigaci{\'o}n Universidad, Junta de Andaluc{\'\i}a and European Regional Development Fund, Project references A-FQM-175-UGR18 and P20-00173 for financial support. We also thank V. Buend{\'\i}a, P.Villegas, R. Corral, J. Pretel, P. Moretti, M. Iba{\~n}ez, P. Garrido, and M. Marsili, for valuable discussions and/or suggestions on earlier versions of the manuscript.
 

\appendix



\def\url#1{}

\bibliographystyle{unsrt}
\bibliography{RG_References.bib}

\end{document}


\maketitle

\section{Extended Methods}

\subsection*{Datasets}
In the experimental setup of Steinmetz {\emph et al.}, mice were exposed to visual stimuli of varying contrast that could appear on the left side, right side, both sides or neither side of a screen. Mice then earned a water reward by turning or holding still a wheel, depending on the type of stimuli. During task performance, two or three Neuropixels probes were inserted at \newd{a} \new{the same} time in the left hemisphere, allowing for high-resolution simultaneous-recordings of hundreds of neurons in several regions during each recording session \cite{Steinmetz}. Thus, for each session the dataset provides the time-stamps and amplitudes of all measured spikes within various brain regions for a total recorded time $T$. In our analysis, we first collected all spikes belonging to a given region, provided it contained more than $128$ simultaneously recorded neurons. Then, an optimal time bin $\Delta t$ was inferred for each area on the basis of the Inter-Spike-Interval distribution of its neurons (see \textit{Time-scale determination} below). This allowed us to convert the spike time-stamps of each neuron into discrete trains of spikes, which were later used to infer pair-wise correlations among neurons and, ultimately, perform RG analyses. 

On the other hand, in the work of Stringer {\emph et al.} mice were presented a sequence of $2800$ images, each remaining for $0.5s$ and intersperse with gray-screen inter-stimuli intervals lasting a random time between $0.3s$ and $1.1s$ \cite{Stringer-Nature}. The resulting task-related activity is thus effectively binned into $\Delta t \approx 0.5s$ time intervals (a time resolution that we keep in our analyses) and two repetitions of the same input-presentation sequence were performed to allow for cv-PCA analyses. Experiments were then repeated with $3$ different mice and using different dimensions of the input ensemble. To construct d-dimensional stimuli, the natural image database was projected onto a set of basis functions that enforced the required dimensionality \cite{Stringer-Nature}. In our analyses we show results for natural, high-dimensional images as well as 4-d, low-dimensional images. 

\subsection*{Time-scale determination}
In order to determine pairwise correlations from empirical neural activity data, it is necessary to discretize the time into bins of certain length $\Delta t$. In our case, to obtain a spike train vector from a list of time stamps $t_{j}^{*}$ at
which a neuron spikes, a common value $\Delta t$ is chosen for all units. In this way, neurons spiking at times $t_{j}$ and $t_{j}+\delta t$ sufficiently close share a bin $\Delta t_{j}$ with non-zero activity. The number of bins for a neuron can be therefore chosen by $T_{i}=\frac{max(t_{j}^{*})}{\Delta t}$, and more generally
we will set $T=max(T_{i})$ so that all neurons have the same amount of bins in their spike trains. 

Choosing the time bin that best transforms the spike times into discrete trains of spikes is an open problem in neuroscience for which several solutions ---such as the use of random bins \cite{tamura_random_2012}, methods that find the best bin size by minimizing a certain cost function \cite{omi_optimizing_2011,Marsili-2019,scholkopf_recipe_2007,ghazizadeh_optimal_2020} and bin-less approaches \cite{victor_binless_2002,paiva_comparison_2010}
have been proposed. The problem becomes even more prominent when working with the simultaneously-recorded activity of a large number of cells as, typically, one finds neurons operating at broadly different time scales within the same area (see Fig.\ref{ISIS_Histogram}), as well as a hierarchy of timescales across regions \cite{kiebel_hierarchy_2008,spitmaan_multiple_2020}.

To perform the RG analyses, we first compute the distribution of inter-spike intervals (ISI) of individual neurons within a particular brain region (Fig.\ref{ISIS_Population}).  Because we are interested in a measure of the "typical" time-scale at which neurons in a population operate, and given also the large neuron-to-neuron variability, we define the optimal time bin $\Delta t$ as the geometric mean of all ISIs from neurons in the population which (as illustrated in Fig.\ref{ISIS_Population}), is a very good estimator of the population characteristic time scale); this geometric-mean value is computed for each region and subsequently used in the corresponding RG analysis (see Table \ref{tab:Table1} for a summary of all areas with their corresponding selected time bin). 

\subsection*{Stationarity of spiking statistics} 
The expression for the maximum eigenvalue $\lambda_{max}$ of the effective connectivity matrix (as derived originally by Dahmen et al. using linear response theory)  relies on the stationarity of the spike trains. Therefore, to assess the stationarity of the recordings in each region we performed an Augmented Dickey-Fuller test (ADF) \cite{Dickey}, which evaluates the null hypothesis that a unit root is present in a time series (i.e., it has some time-dependent structure and it is therefore non-stationary). The alternative hypothesis, should the null hypothesis be rejected, is that the time series is stationary. The \textit{adfuller} function from the \textit{statsmodel} Python library was used to perform the analysis.

Thus, for each neuron in a particular region and experiment we tested the null hypothesis at a 5\% significance level, then counted the fraction of neurons in the experiment for which the null hypothesis could not be rejected (i.e. had non-stationary activity). This fraction was then averaged over all experiments to estimate the percentage of neurons with non-stationary activity for a given region (see Table \ref{tab:Table1}), and only those experiments with less than 10\% of the neurons showing non-stationary spiking statistics were considered in the subsequent analyses.

\subsection*{Autocorrelation decay times}
Spike-trains were first smoothed using a Gaussian-kernel with $\sigma=50ms$ to obtain a firing rate time-series. For each neuron, the autocorrelation function was then computed, averaging across all neurons to obtain an estimate of the autocorrelation for the population (reported for the MOp region as an example in Fig.4A of the main text). We then fitted the resulting decaying function to an exponential, extracting a characteristic time-scale $\tau_{Corr}$ for each experiment. Finally, these values were averaged across experiments to obtain the mean autocorrelation decay time in each region (see Table \ref{tab:Table1}). The results allow us to define an optimal $\Delta t=1s$ over which to compute spike-counts covariances, sufficient for autocorrelations to decay (a requirement for the equivalence between integrated time-lagged and spike-count covariances to hold) while maximizing the amount of available samples.

\subsection*{Likelihood ratio test for powerlaw distribution of eigenvalues}

In the limit of $N\rightarrow\infty$ there exist a direct relationship between the exponent of the rank-ordered eigenvalues and their probability density (see Section 5 below). Therefore, we leveraged the approach in \cite{Clauset} to test whether a power-law distribution is indeed the best description of the density of eigenvalues for the different regions in the data set. Following \cite{Clauset} we computed the loglikelihood ratio $R$ between the estimated power-law and equivalent exponential and lognormal candidate distributions, together with the the p-value for the significance of the test. In 14 out of the 16 regions, the exponential distribution provides a worse fit ---tested at a 5\% significance level--- than the corresponding power-law ($R>0$ ; $p\leq0.05$). On the other hand, the lognormal distribution always provides an equally good description of the eigenvalue density ($p>0.05$), and cannot be ruled out as the underlying distribution. These values, together with the R-squared value for the best powerlaw and exponential fits of the rank-ordered eigenvalues in each region, are listed in Table \ref{tab:Table2}, whereas the corresponding fitted curves are plotted in Fig.\ref{fig:Exp_vs_Pow}.

\subsection*{Estimation of parameter g}
The dynamics of a neuron in a recurrent network model of linear units driven by noise is described by the following equation:
\begin{equation}
     \dot{x_i}(t)=-x_i(t)+ g \sum_{i=1}^N J_{ij} x_j +\xi_i(t),\:\:\:\:i=1,...,N
    \label{eq:rate}
\end{equation}
with i.i.d. Gaussian connectivity $J_{ij}\sim \mathcal{N}(0,1/N)$ and $\langle\xi_i(t) \xi_j(t+\phi)\rangle=\delta_{ij}\delta(\phi)$. For this type of model, the dynamics becomes unstable at a critical value $g_c=1$ \cite{YuHu}, where $g$ describes the overall connection strength (i.e., the spectral radius or maximum eigenvalue of the connectivity matrix, $\lambda_{max}$, as defined in \cite{Helias}) and can be considered a control parameter of the system. In the large $N$ limit the analytical form of the probability density of the covariance eigenvalues has been derived in \cite{YuHu}, and reads:
\begin{equation}
 p_{RC}(x)=\frac{3^\frac{1}{6}}{2\pi g^2 x^2}\left[\sum_{\xi=1,-1} \xi\left( (1+\frac{g^2}{2})x-\frac{1}{9}+\xi \sqrt{\frac{(1-g^2)^3 x(x_+-x)(x-x_-)}{3}} \right)^\frac{1}{3} \right], \:\: x_-\leq x \leq x_+,
\end{equation}
with
\begin{equation}
    x_{\pm}=\frac{2+5g^2-\frac{g^4}{4}\pm\frac{1}{4}g(8+g^2)^\frac{3}{2}}.
\end{equation}

Thus, given a data set of simultaneously recorded neurons, one can infer the dynamical regime of the population activity by fitting the sampled covariance matrix eigenvalue spectrum to the theoretical distribution. More concretely, we looked for the value of $g$ that minimized the $L^2$ error using the Cramer-von Mises statistics between the empirical cumulative distribution $F_n(x)$ and the theoretical one $F(x)=\int_{-\infty}^x p_{RC}(x)dx$ \cite{YuHu}:
\begin{equation}
    D^2_{CvM}=\int (F(x)-F_n(x))^2 dF_n(x)=\frac{1}{12n^2}+\frac{1}{n}\sum_{i=1}^n \left(F(x_i)-\frac{2i-1}{2n} \right),
\end{equation}
where $n$ is the total number of samples and $x_i$ are the eigenvalues of the empirical long time window covariance matrix.
For a given region, we computed $D^2_{CvM}$ for each experiment, then averaged across all experiments (values are provided in Table \ref{tab:Table1}).

\section{Further results of the RG analyses}

Here we provide extended results illustrating the ubiquity of scaling across brain regions. 

\subsection*{Non-zero activity distributions across brain regions}
\begin{figure}
\centering
\includegraphics[width=14.0cm]{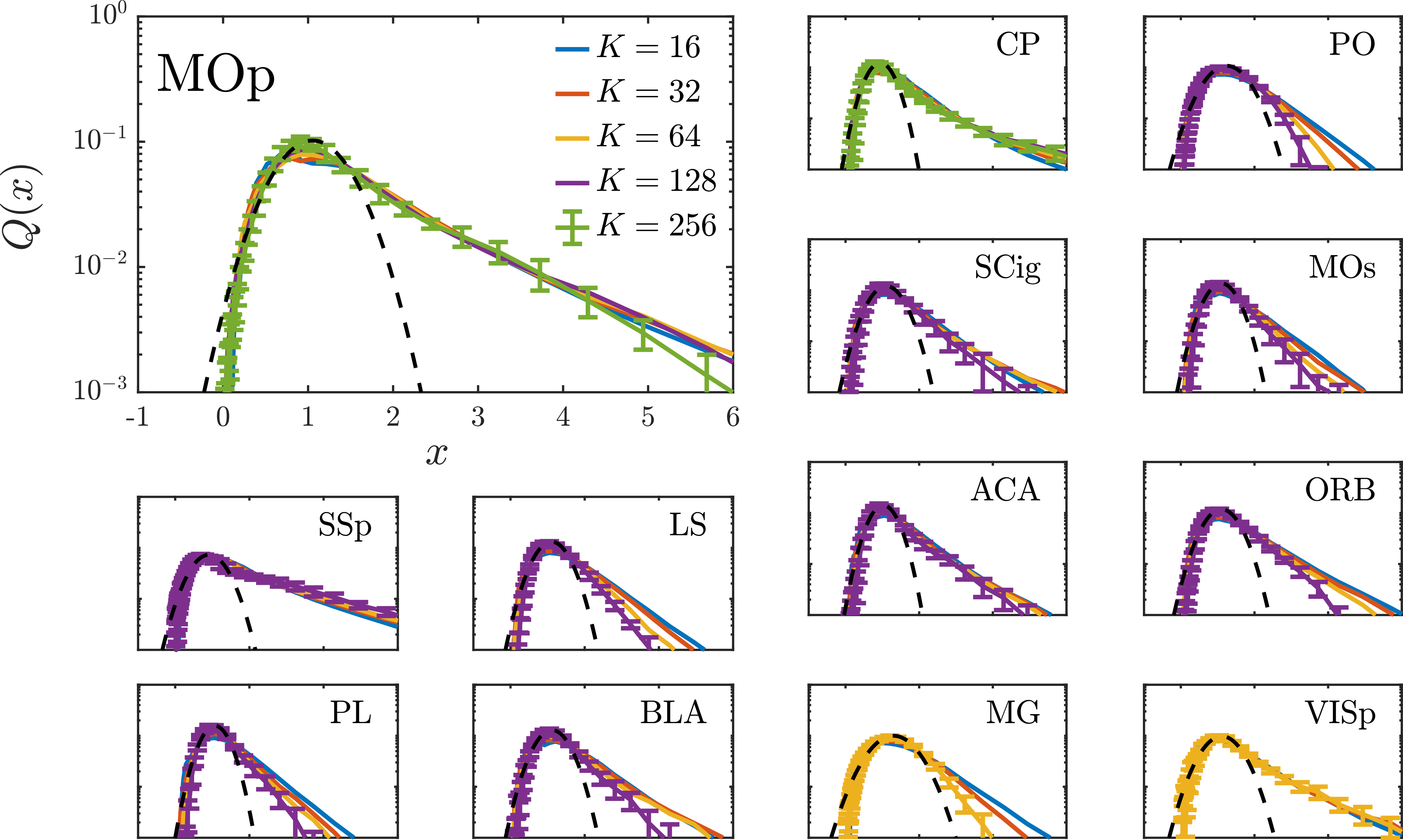}
\caption{Non-zero activity distribution curves at different levels of coarse-graining in $13$ different areas of the mouse brain as presented in the main text. Significant tails towards large values of the activity are observed, while curves are fairly invariant in most areas across RG steps, suggesting the existence of a non-Gaussian fixed point of the RG flow. Errors (shown only for the curve corresponding to the last step of coarse-graining) are computed as the standard deviation over random quarters of the data. \label{Qx_vs_rgsteps}}
\end{figure}

Fig.\ref{Qx_vs_rgsteps} shows the normalized probability distribution function for the non-zero activity inside clusters of size $K$ (and averaged over all existent clusters) at $5$ different steps of the coarse-graining. As in the main text, errors are computed as the standard deviation over random quarters of the data, and only shown in the last step of coarse-graining.

\subsection*{Scaling of time-decaying auto-correlation curves}
\begin{figure}
\centering
\includegraphics[width=14cm]{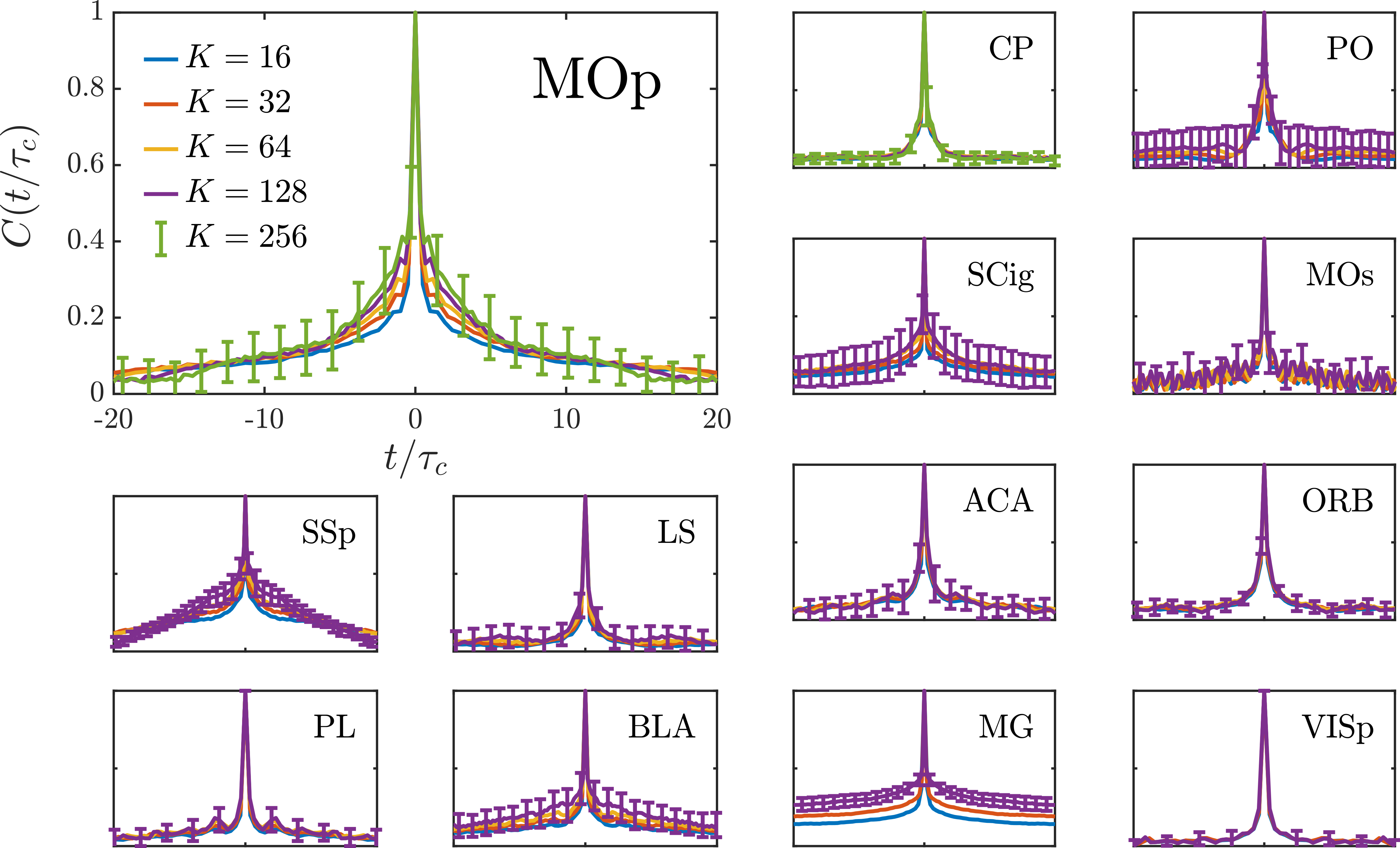}
\caption{Mean normalized correlation function of coarse-grained variables during the RG flow, measured in 13 different areas of the mouse brain. Time is re-scaled for each curve by the characteristic time scale $\tau_{c}(K)$, computed as the $1/e$ point of the decay. Errors shown in the last step of coarse-graining are computed as the standard deviation over random quarters of the data \label{Correlation_curves}}
\end{figure}

\begin{figure}
\centering
\includegraphics[width=14cm]{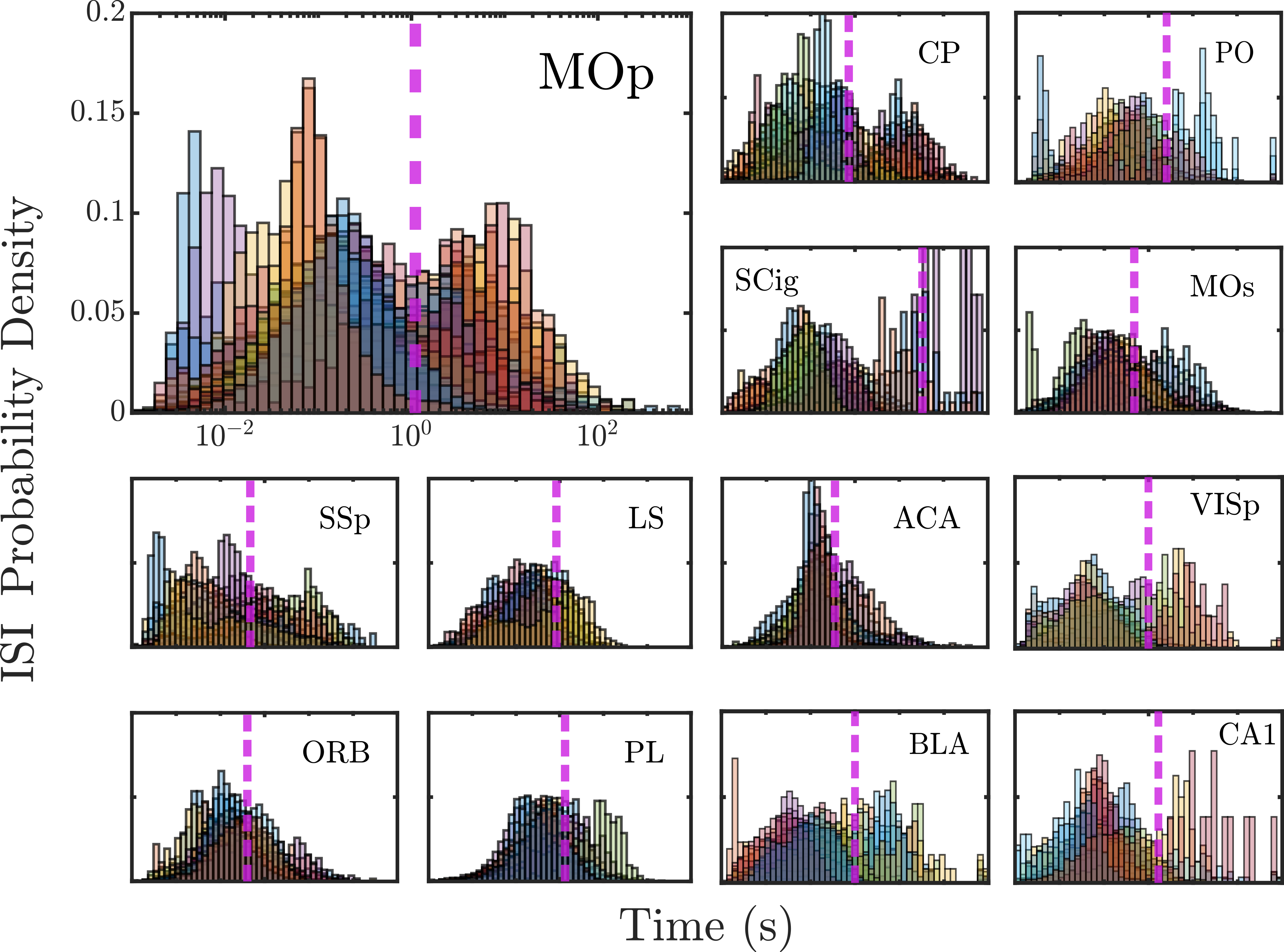}
\caption{Histograms estimating the probability of finding a certain value of the inter-spike interval (ISI) in the resting-state activity of individual neurons belonging to a particular area
(shown, as before, for 13 areas). Each color represents the distribution for a single neuron (only a few neurons are shown
for the sake of clarity), while pink dashed lines mark the geometric mean ISI for each area.
Let us remark that the arithmetic mean does not properly represent 
the characteristic time scale
in each region.
\label{ISIS_Histogram}}
\end{figure}

\begin{figure} 
\includegraphics[width=14cm]{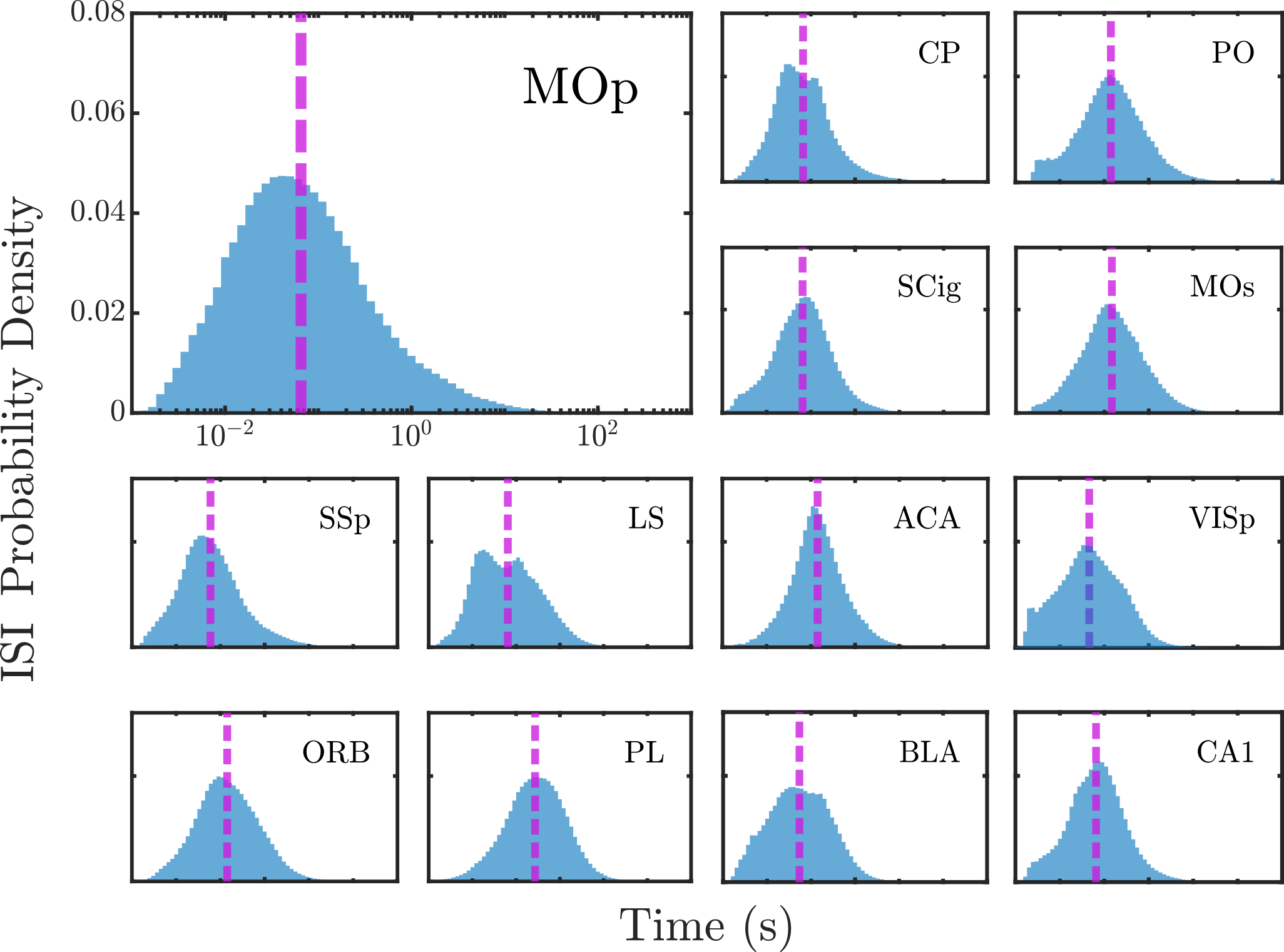}
\caption{Individual-neuron inter-spike interval distributions for $13$ different regions of the mouse brain as determined using all recorded neurons in each population (for each of them, there is a large neuron-to-neuron variability as shown in Fig.\ref{ISIS_Histogram}). The characteristic or  "optimal" time bin is defined as the geometric mean of the ISI distribution (vertical dashed lines).}\label{ISIS_Population}
\end{figure}

 Fig.\ref{Correlation_curves} shows 
the auto-correlation function for resting-state type of activity, computed as:
\begin{equation}
C^{(k)}(t)=\dfrac{1}{N_{k}}\sum_{i=1}^{N_{k}}\frac{\langle  x_{i}^{(k)}(t_{0})x_{i}^{(k)}(t_{0}+t)\rangle - \langle x_{i}^{(k)}\rangle^{2}}
{\langle\left(x_{i}^{(k)}\right)^{2}\rangle - \langle x_{i}^{(k)}\rangle^{2}}
\end{equation}
and normalized to lie within the unit interval.
In each subplot, the above quantity is plotted against the re-scaled time $t/\tau_{c}$, where $\tau_{c}$ is the characteristic time of the exponential decay, for a particular region of the mouse brain at for $4$ different levels of coarse-graining.  For ease of visualization, the errors ---computed as the standard deviation over random (non-shuffled) quarters of the data--- are only shown in the last step of coarse-graining procedure.

\subsection*{Robustness against time bin choice}\label{SI_Robustness}
Activity of neurons, even within the same brain region, can span many time scales, with some neurons typically firing in the range of milliseconds and with a characteristic time-scale of the order of seconds (see Fig.\ref{ISIS_Histogram}). Thus, in selecting a common time bin to convert the spike times of all neurons within a region into discrete trains of spikes, we irretrievably lose some information.

To assess whether the exponent values reported in the main text are robust against our choice of $\Delta t$, we repeated the same RG analyses using different time-windows $\Delta t \in [0.01s, 4s]$ to bin the resting-state activity of each region. 

\begin{figure}
\centering  
\includegraphics[width=14cm]{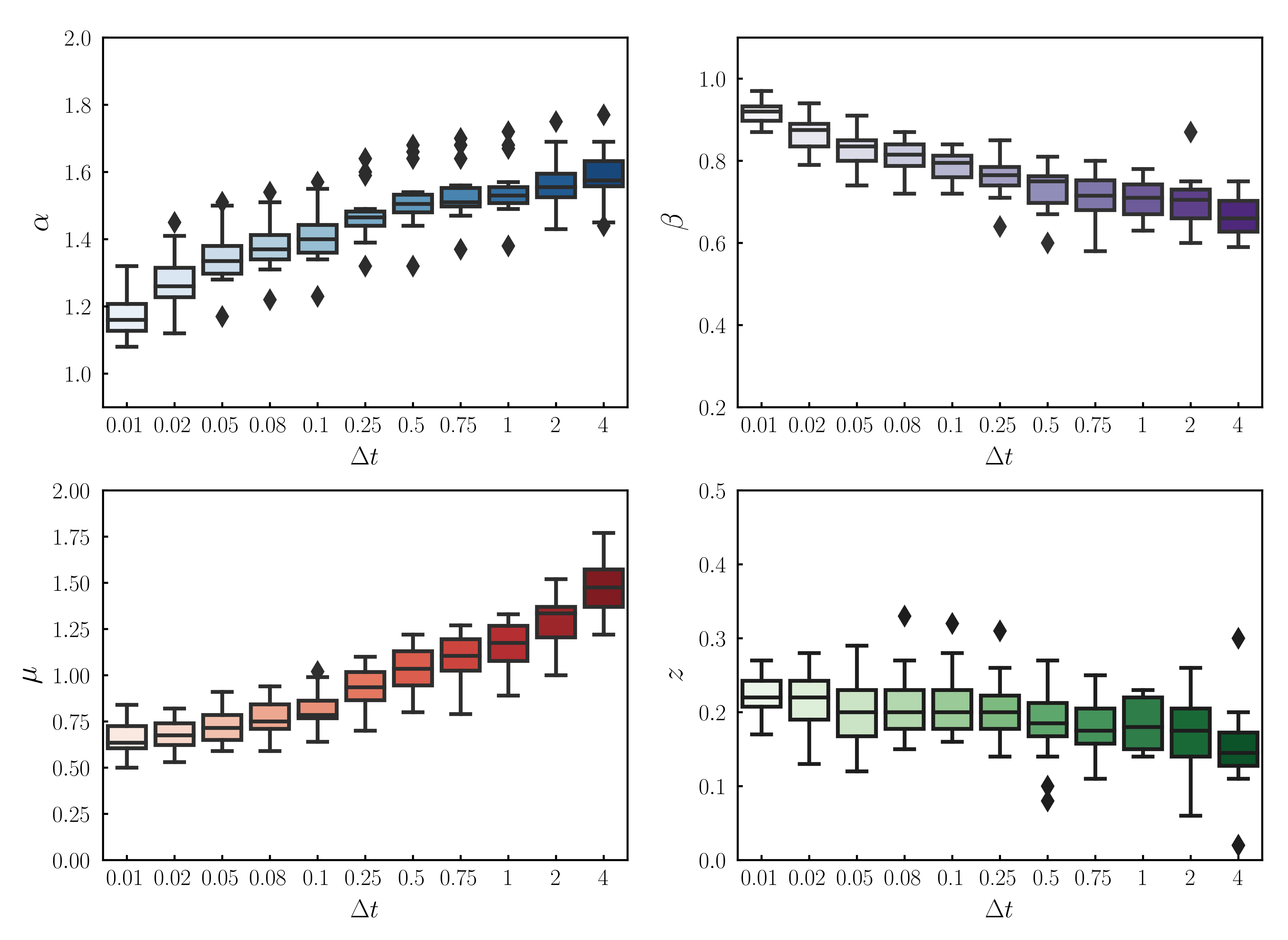}
\caption{Change of average scaling exponents as a function of the employed time bin within the range $\Delta t \in [0.01s, 4s]$ for each region. The line inside of each box represents the sample median, the whiskers reach the non-outlier maximum and minimum values;  the distance between the top (upper quartile) and bottom (lower quartile) edges of each box is the inter-quartile range (IQR). Black diamonds represent outliers, i.e. values that are more than $1.5 IQR$ away from the limits of the box. \label{Exponents_vs_bins}}
\end{figure}

As illustrated in Fig.\ref{Exponents_vs_bins}, the values of the exponents $\alpha$, $\beta$ and $z$ turn out to be fairly robust within a broad range of biologically plausible time scales. As one could trivially expect, broader time bins increase the probability of finding neurons spiking simultaneously, i.e. appearing as more strongly correlated. This, in turn, has the effect of shifting the values of the exponents $\alpha$ and $\beta$ for the variance and free-energy scalings slightly towards those expected for a fully correlated scenario. On the other hand, the exponent $\mu$ remains fairly unchanged for binning times up to $\sim 100ms$, but then increases on longer time scales beyond the typical geometric mean ISI of the regions.  

\subsection*{Resting-state vs task-induced activity}
\begin{figure}
\centering
\includegraphics[width=14cm]{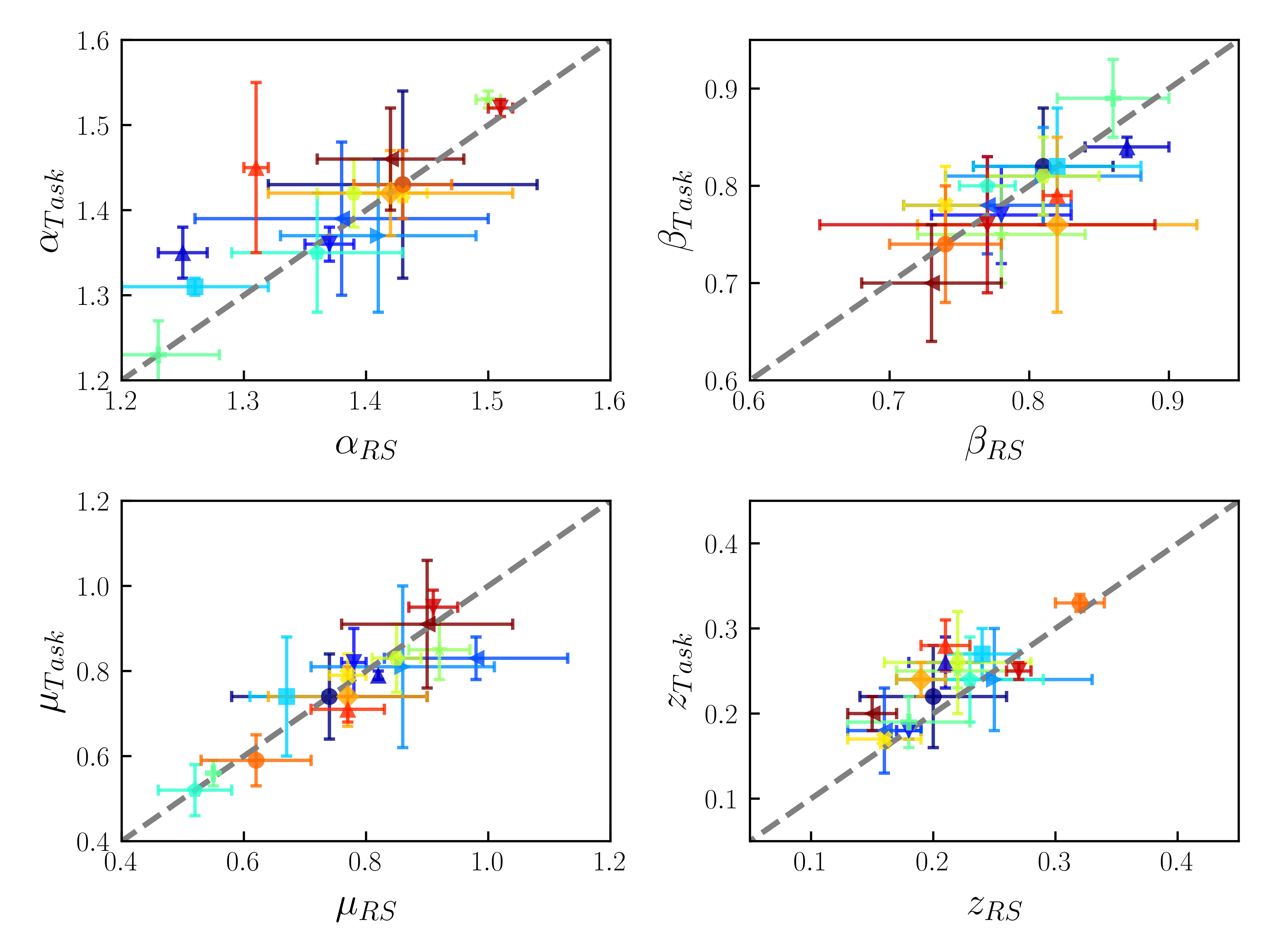}
\caption{Scaling exponents for task-induced vs resting-state type of activity in 16 different regions of the mouse brain. Markers and colors for each area have been chosen as defined by Fig.1 in the main text. Error bars are defined as the standard deviation of the corresponding quantity across all experiments belonging to the same region. \label{Spont_vs_Input}}
\end{figure}

We plotted in Fig.\ref{Spont_vs_Input} the four measured scaling exponents in task-induced vs resting-state type of activity, with results that clearly support the idea that the observed exponents are almost invariant with the mouse behavioral state. Only small deviations are observed in a few regions.

\subsection*{Non-universal $\alpha$ exponent in input-encoding subspace}
\begin{figure}
\centering
\includegraphics[width=14cm]{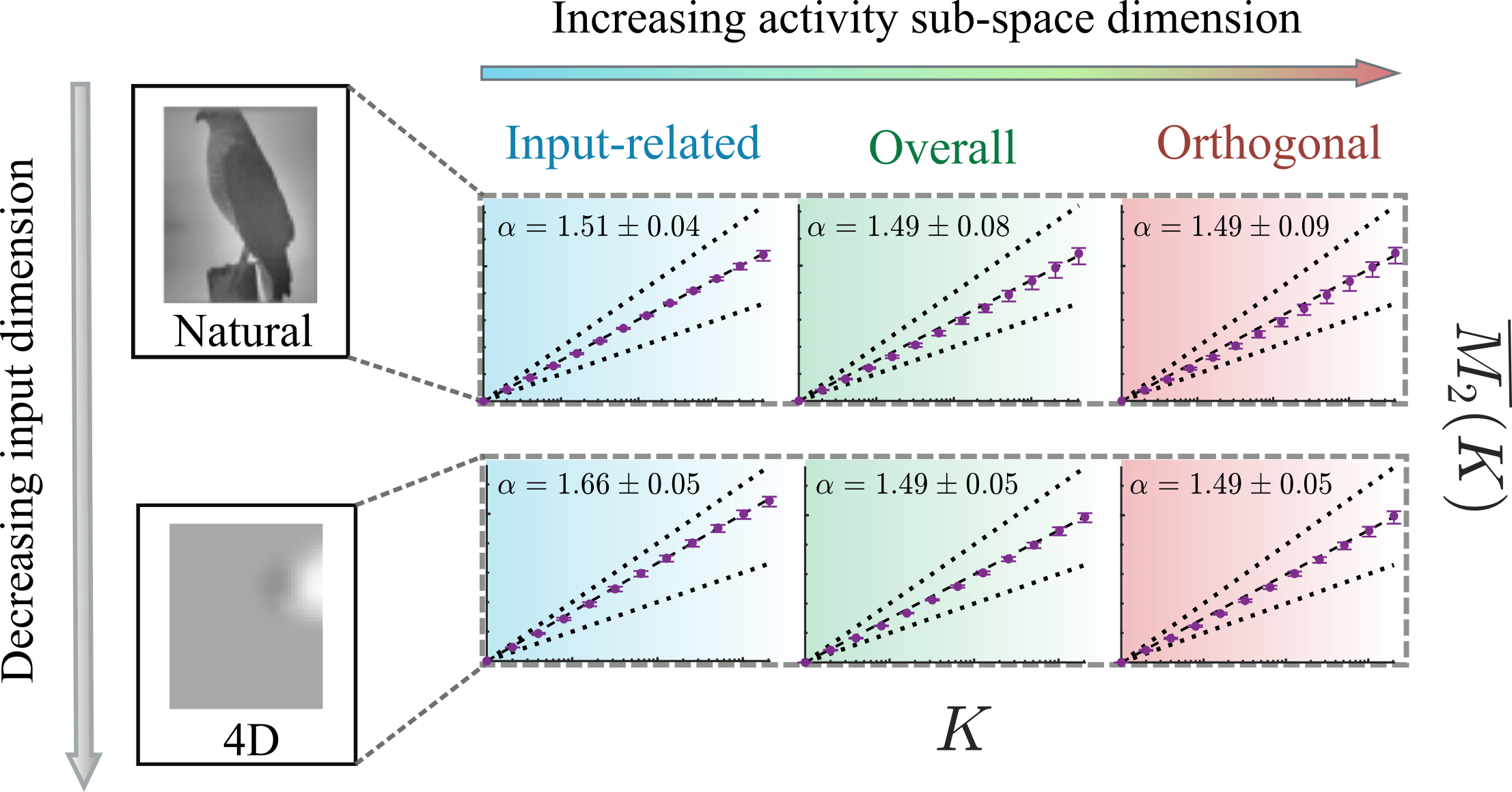}
\caption{Diagram showing the two observed trends in the power-law exponent $\alpha$, which characterizes the scaling of the variance as a function of their rank. $\alpha$ increases with the complexity of the input when activity is projected into the task-encoding subspace (also called representation manifold). On the other hand, for a fixed type of input, the encoding activity shows a higher exponent (meaning higher correlations), while the background activity ---which lies in a higher-dimensional subspace orthogonal to the representational manifold--- and the overall/raw activity share the same exponents. Natural and low-dimensional images examples have been adapted from \cite{Stringer-Nature}.
\label{Alpha_FULL}}
\end{figure}

Fig.\ref{Alpha_FULL}
shows the scaling exponent for the variance of the activity inside block-neurons, measured over the overall raw data, the input-related/encoding activity, and the background activity. We only observe a significant change on the exponent ---for both types of stimuli: natural and 4D images--- when the input-encoding activity is considered, but not in the background, orthogonal subspace. Observe also, that there is no significant change in the $\alpha$ exponent in orthogonal components as compared to raw/overall activity.

\subsection*{Dependence of $\mu$ with the samples-to-neurons ratio}

The spectrum of covariance eigenvalues is known to be strongly sensitive to the ratio $T/N$, where $N$ is the total number of neurons recorded in the experiment and $T$ is the number of time bins \cite{kong_spectrum_2017}. 

\begin{figure}
\centering
\includegraphics[width=14cm]{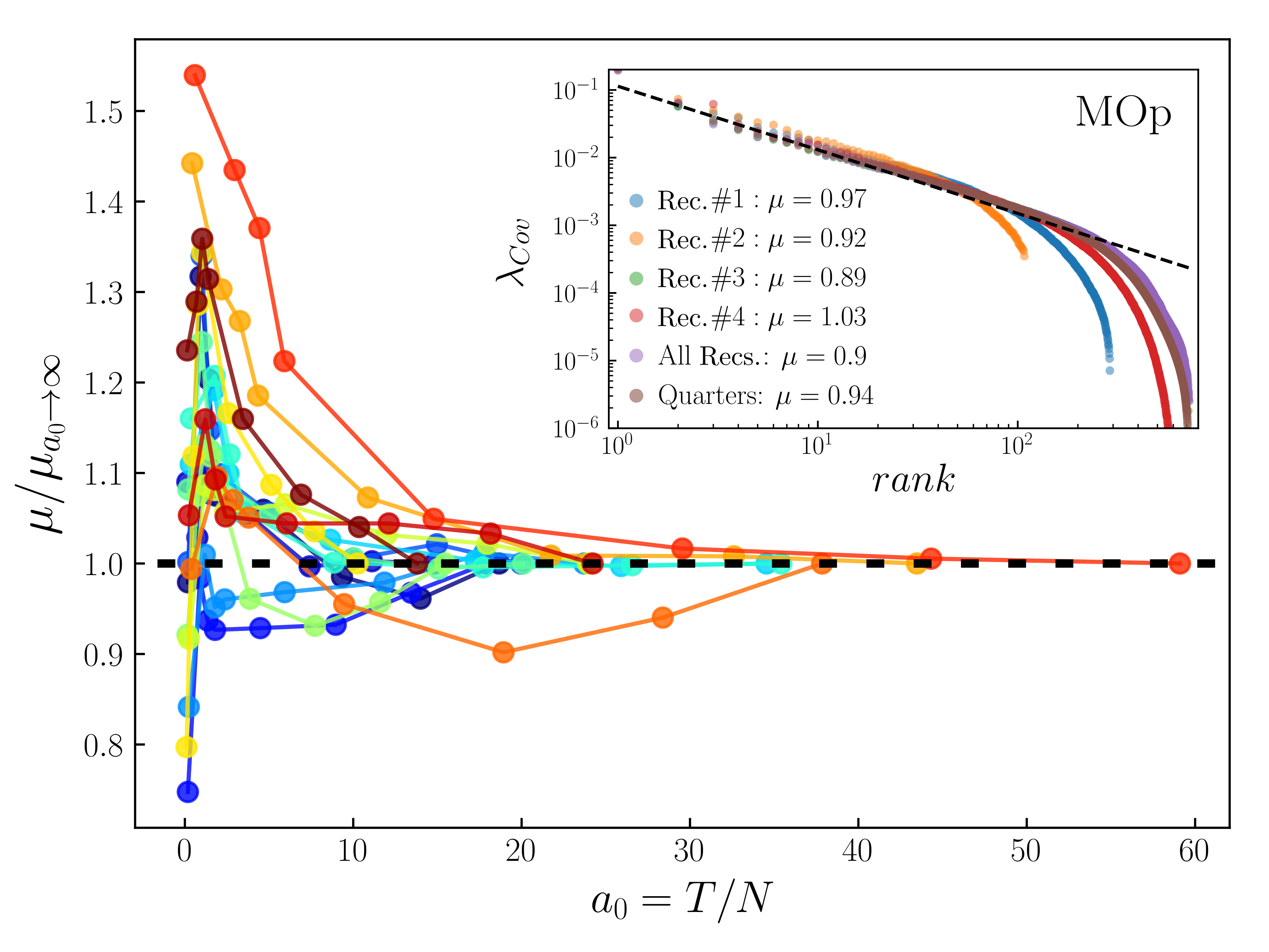}
\caption{For each region, the scaling exponent for the covariance matrix eigenvalues at the last step of RG is plotted against the ratio samples-to-neurons  in sub-sampled experiments, then re-scaled by its expected value when $a_0 \rightarrow \infty$ (estimated by using the full length of the time series). For each region, the experiment with the greatest number of recorded neurons was considered. \textbf{Inset:} Rank-ordered plot of eigenvalues in the MOp region using all the neurons while taking (i) each of the four recordings belonging to a experiment separately; (ii) merging them together; and (iii) averaging over quarter-splits of the data. Line fit is over quarters of the data.}
\label{fig:Muvsa0}
\end{figure}

\begin{figure}
\centering
\includegraphics[width=14cm]{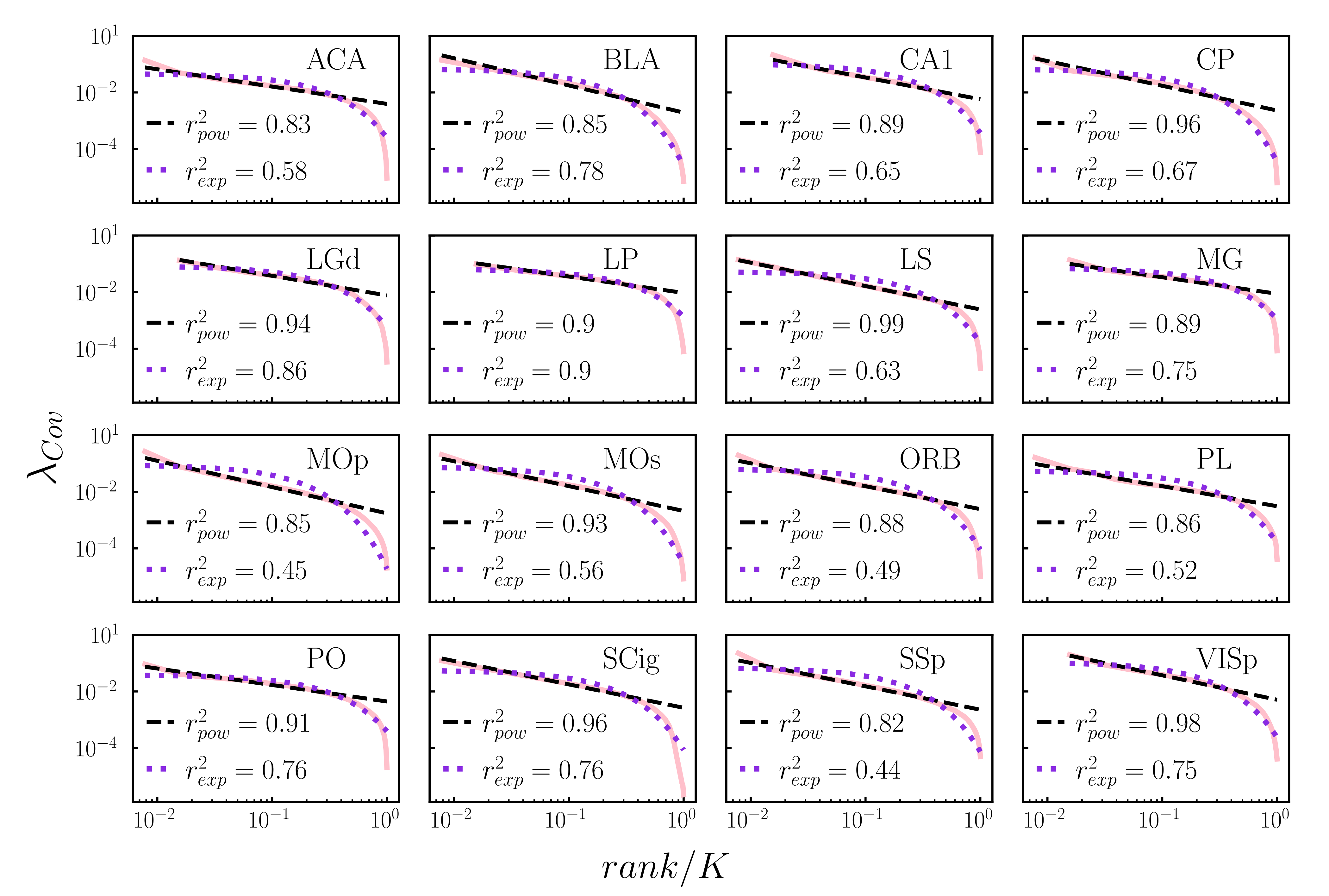}
\caption{Best power-law (black dashed line) and exponential (purple dotted line) fits for the covariance matrix eigenvalues at the last step of RG (pink line), together with their corresponding R-Squared values. For each region, the experiment with the greatest number of recorded neurons was considered.}
\label{fig:Exp_vs_Pow}
\end{figure}

In Fig.\ref{fig:Muvsa0} we scrutinize the dependence of the scaling exponent for the rank-ordered eigenvalues of the covariance matrix $\mu$ with the ratio $a_{0} = T/N$. The figure shows that the estimates converge to the asymptotic value only for sufficiently large values $a_{0} > 20$, while $\mu$ typically increases as one moves towards the sub-sampled regime. 

In the inset of Fig.\ref{fig:Muvsa0} we compare the fitted exponents for the rank-ordered plot of eigenvalues in the example region MOp when i) taking the activity recorded from each of the 4 intervals of spontaneous activity in the recording session, ii) merging the 4 intervals together and iii) averaging over quarter splits of the data set where the intervals were merged together.
In individual spontaneous intervals $a_{0} > 20$ is not always fulfilled, thus causing variability in the fitted exponents. By merging together all recordings of spontaneous activity belonging to the same experiment we can typically ensure that $a_{0} > 20$, even if analyses are performed over quarter-splits of data. This confirms the consistency and robustness of the estimated exponents.

\section{RG analyses on surrogated data}
As a sanity check, we performed the RG analysis over surrogated data in three different ways: i) by randomly shuffling all the spikes for each neuron; ii) by shifting all the spikes of each neuron a random amount, so that the structure within the spike train is preserved; and iii) by randomizing the spikes across neurons but not time. Here, we chose the MOp region for having the greatest number of recorded neurons, and compare the results with the same analysis in the unsurrogated case (Fig.\ref{unsurrogated})

We observe that time series shuffle (Fig.\ref{shuffledspikes}) or shifting (Fig.\ref{shiftedspikes}) destroys the non-trivial scaling properties, showing no signs of convergence towards a non-Gaussian fixed point of the probability distribution, and exponents $\alpha$, $\beta$ and $z$ akin to those expected for an independent-neurons model. The power-law spectrum of the clusters covariance matrix still seems to decay with a smaller, non-trivial exponent, probably due to the firing rate heterogeneity in the population activity, but closer inspection reveals that a power-law dependence is no longer suitable when compared to the unsurrogated case.

\section{RG analyses on a randomly recurrent network of linear rate neurons}
\begin{figure}
\centering
\includegraphics[width=14cm]{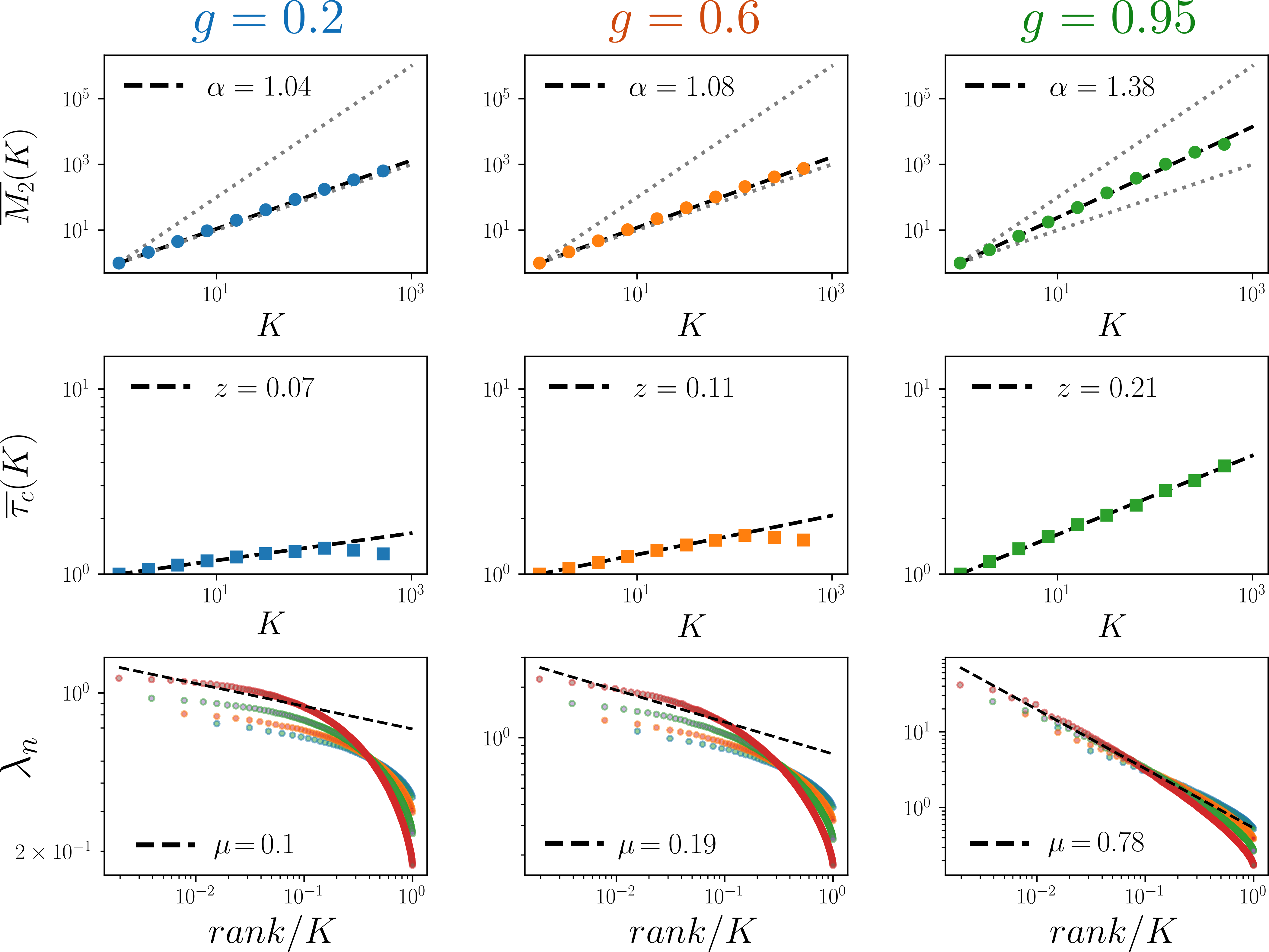}
\caption{RG over randomly recurrent network of linear rate neurons for different values of the overall coupling strength $g$. Top: Variance of the non-normalized activity, Middle: scaling of the characteristic correlation time, Bottom: Covariance matrix spectrum for the resting-state activity in
clusters of size $K$, as a function of the block-neuron size. Systems closer to the edge-of-instability ($g_c=1$) show power-law scalings similar to the ones observed in the data. Neuron blocks in systems further away from such a regime behave like uncorrelated variables. In addition the rank-ordered spectrum is flatter and the scaling across RG steps is lost.}
\label{RGYuHu}
\end{figure}

\begin{figure}
\centering
\includegraphics[width=14cm]{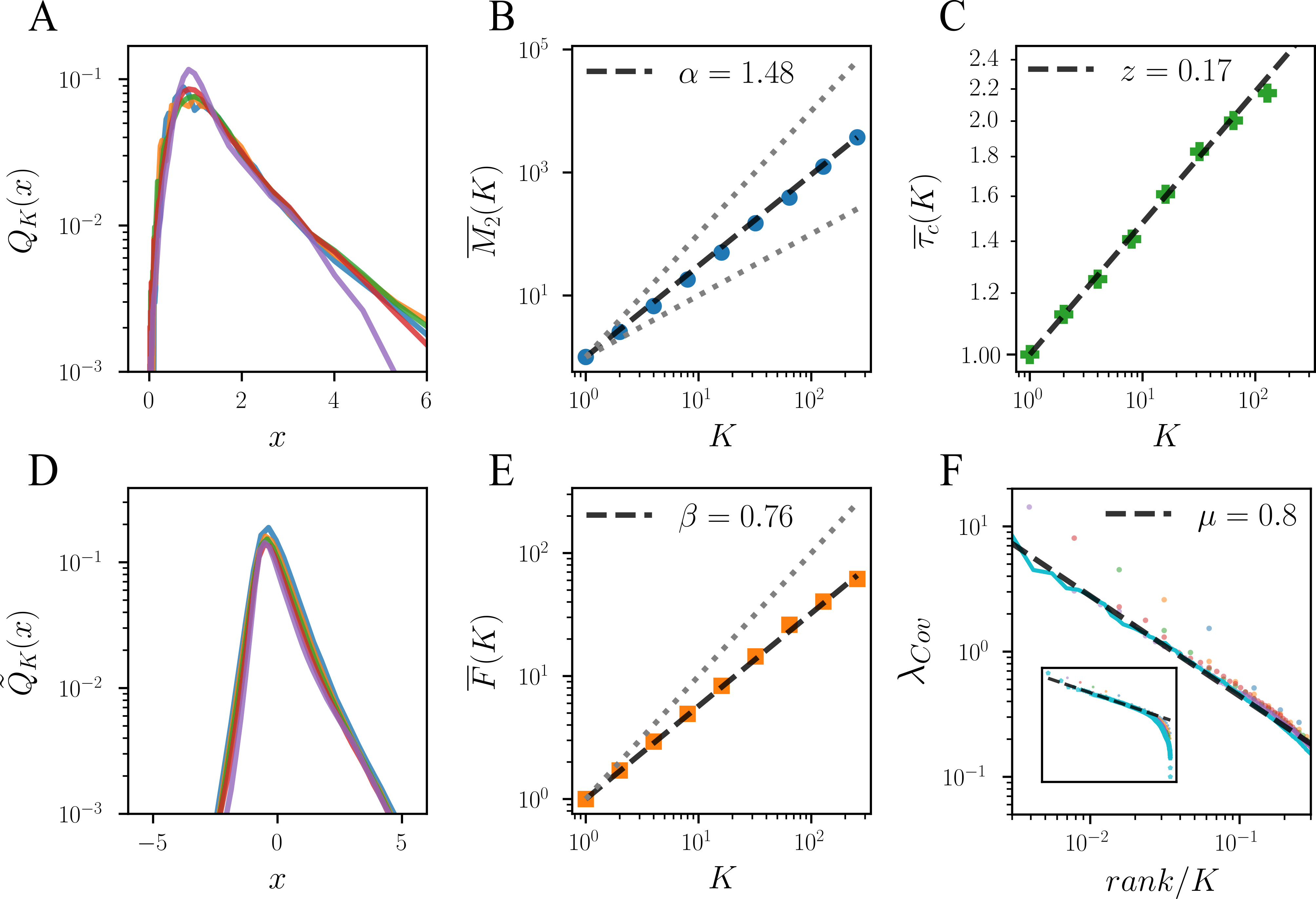}
\caption{RG analysis over non-surrogated (control) data, for the region MOp. Plotted for comparison with surrogated reshuffled sets in Fig.\ref{shuffledspikes}, \ref{shiftedspikes} and \ref{shuffledneurons}. (A) Probability distribution for normalized non-zero activity in clusters of size $K=\{16, 32, 64, 128, 256\}$, corresponding to the yellow, blue, green, red and purple curves, respectively. (B) Scaling of activity variance inside clusters with cluster size. (C) Scaling of characteristic autocorrelation time with cluster size. (D) Probability distribution for normalized activity in momentum space, setting a cutoff on the first $K=\{16, 32, 64, 128, 256\}$ eigenmodes (computed as in \cite{Bialek-largo}, color code as in panel A). (E) Scaling of the "free-energy" with cluster size. (F) Inset: Covariance matrix spectrum with power law fit against the normalized rank. Main: closer view showing small-rank outliers.}
\label{unsurrogated}
\end{figure}

\begin{figure}
\centering
\includegraphics[width=14cm]{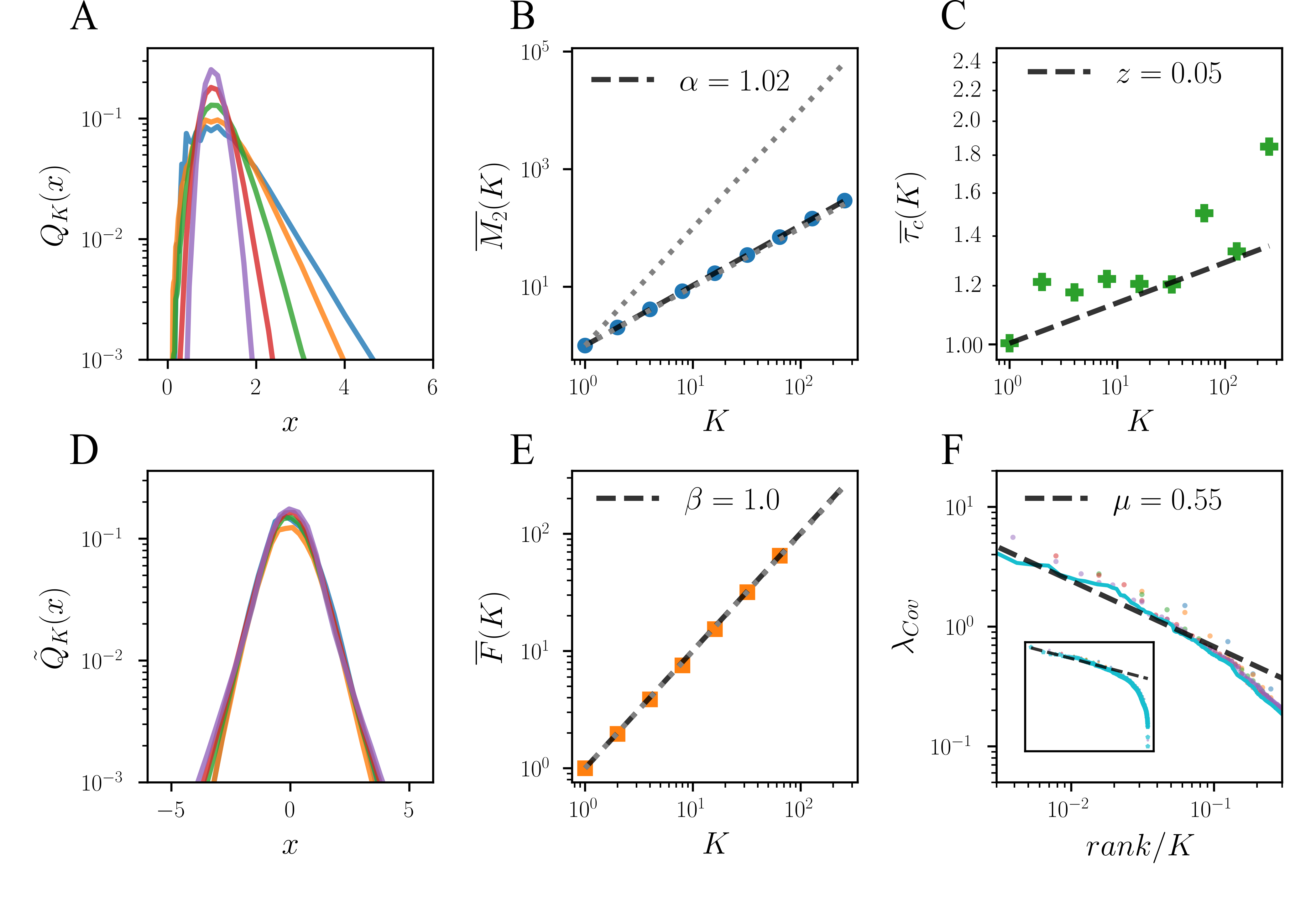}
\caption{RG analysis over surrogated data created by  by randomly shuffling all the spikes for each neuron. (A) Probability distribution for normalized non-zero activity in clusters of size $K=\{16, 32, 64, 128, 256\}$, corresponding to the yellow, blue, green, red and purple curves, respectively. (B) Scaling of activity variance inside clusters with cluster size. (C) Scaling of characteristic autocorrelation time with cluster size. (D) Probability distribution for normalized activity in momentum space, setting a cutoff on the first $K=\{16, 32, 64, 128, 256\}$ eigenmodes (computed as in \cite{Bialek-largo}, color code as in panel A) (E) Scaling of the "free-energy" with cluster size. (F) Inset: Covariance matrix spectrum with power law fit against the normalized rank. Main: closer view}
\label{shuffledspikes}
\end{figure}

\begin{figure}
\centering
\includegraphics[width=14cm]{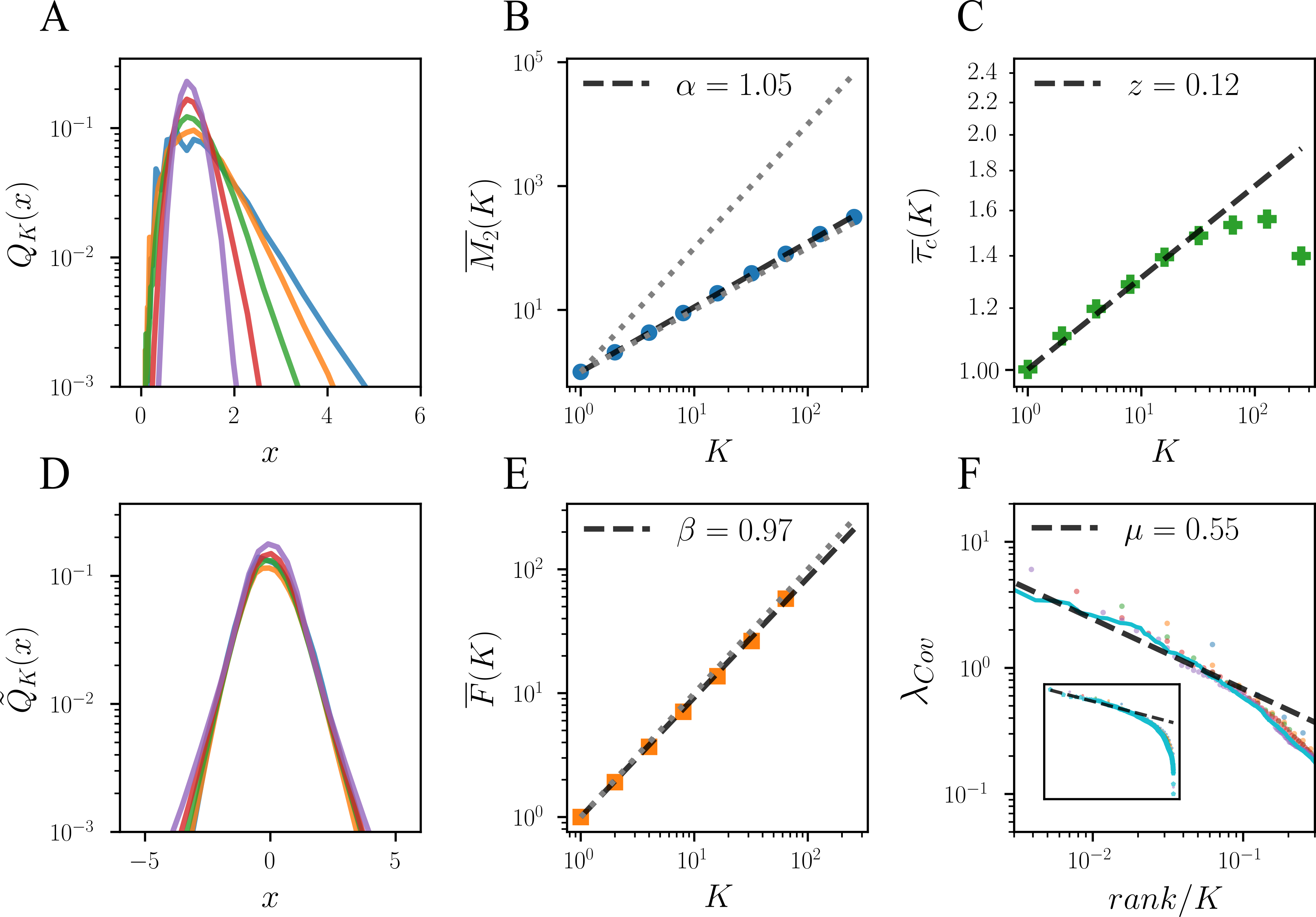}
\caption{RG analysis over surrogated data created by shifting all the spikes of each neuron a random amount, so that the structure within the spike train is preserved. (A) Probability distribution for normalized non-zero activity in clusters of size $K=\{16, 32, 64, 128, 256\}$, corresponding to the yellow, blue, green, red and purple curves, respectively. (B) Scaling of activity variance inside clusters with cluster size. (C) Scaling of characteristic autocorrelation time with cluster size. (D) Probability distribution for normalized activity in momentum space, setting a cutoff on the first $K=\{16, 32, 64, 128, 256\}$ eigenmodes (computed as in \cite{Bialek-largo}, color code as in panel A) (E) Scaling of the "free-energy" with cluster size. (F) Inset: Covariance matrix spectrum with power law fit against the normalized rank. Main: closer view.}
\label{shiftedspikes}
\end{figure}

We simulate a randomly connected recurrent network of linear rate neurons driven by noise described by Eq.\ref{eq:rate}. The simulations were ran for networks of size $N=1024$, choosing an integration step $h=0.01$ while setting $\sigma=1$. We then binned the resulting time series with a $\Delta t=0.1$ sampling window, and performed the RG analyses for various values of the control parameter $g$ (Fig.\ref{RGYuHu}). In particular, we show the scaling of the variance, autocorrelation time, and covariance eigenvalues inside the clusters formed along the RG flow. We observe that power-law scaling (with exponent values $\alpha$ and $z$ similar to the empirically measured ones) can be retrieved only when the system is close to the edge-of-instability ($g=0.95$). Away from such a regime one finds the trivial expectations for asymptotically uncorrelated variables ($\alpha=1$, $z \approx 0$).
As for the covariance-matrix spectrum, we see an increasing overlap of the rank-ordered spectrum across RG steps and a tendency towards scaling collapse and power-law-like behavior as the parameters gets closer to criticality. 
Since the activity of the neurons in the model take positive and negative values within a continuous range, the exponent $\beta$ for the scaling of the probability of being silent is trivially zero (i.e. this property cannot be described by the simple linear model).

\section{Rank-ordering and spectral density of covariance matrix eigenvalues}\label{rank_vs_density}

Let us provide a simple mathematical calculation that highlights the relationship between the power-law exponents for the probability distribution of eigenvalues, that we also call "spectral density", and eigenvalue-vs-rank plots. 

Let us assume that one has measured the spectrum of a covariance matrix extracted from some experimental data and found that the $n$-th largest eigenvalue scales with its rank (when ordered from  the largest to the smallest) as:
\begin{equation}
\lambda_{n}=An^{-\gamma}.
\end{equation}
Given such a dependency, what can be said about the spectral density $\rho(\lambda)$, which describes the probability of finding a certain value of $\lambda$ in the spectrum? Formally, and since
there is a discrete number of eigenvalues (limited by the number of
variables $N$):
\begin{equation}
\rho(\lambda)=\frac{1}{N}\sum_{n}\delta\left(\lambda-\lambda_{n}\right).
\end{equation}
Nevertheless, in the limit of $N\rightarrow\infty$ one can approximate
the above distribution by a continuous one:
\begin{equation}
\rho(\lambda)\approx\frac{1}{N}\int_{1}^{\infty}\delta\left(\lambda-\lambda_{x}\right)dx.
\end{equation}
Let us now define the function $g(x)=\lambda-\lambda_{x}=\lambda-Ax^{-\gamma}$,
which has a root at $x_{0}=\left(\lambda/A\right)^{-1/\gamma}$. Using
the properties of the Dirac delta distribution:
\begin{equation}
\delta\left(g(x)\right)=\dfrac{\delta\left(x-x_{0}\right)}{\left|g'\left(x_{0}\right)\right|}=\dfrac{\delta\left(x-x_{0}\right)}{\nu A\left(\frac{\lambda}{A}\right)^{1+1/\gamma}}
\end{equation}
one can then, finally, write the spectral density in continuous approximation
as:
\begin{equation}
\rho(\lambda)\approx\frac{1}{N}\left(\nu A\left(\frac{\lambda}{A}\right)^{1+1/\gamma}\right)^{-1}\int_{1}^{\infty}\delta\left(x-x_{0}\right)dx=\frac{A^{1/\nu}}{N\nu}\lambda^{-1-1/\gamma}
\end{equation}
Thus, we can conclude that given a power-law dependency $\lambda_n \sim n^{-\gamma}$ of the eigenvalues on their rank,  its corresponding spectral density is expected to  also follow a power-law $\rho(\lambda) \sim \lambda ^{-\nu}$, with $\nu = 1 + 1/\gamma$.  For a more detailed discussion of this scaling relation see \cite{Andrea,Li-Zipf}. 


\section{cvPCA: noise vs stimuli-related activity}\label{SI_cvPCA}

Following Stringer {\emph{et al.}}, let us consider two observation matrices, $X_{(1)}, X_{(2)}\in\mathbb{R}^{N\times T}$,
corresponding to two identical realizations or trials of the experiment, with $N$ the number of neurons and $T$ the number of images or stimuli presented. Thus, we denote by $x_{ij}^{(k)}$ the matrix element containing the activity of neuron $i$ during presentation of image $j$ in trial $k$. For simplicity, we will further assume that the mean activity of each neuron across images has been subtracted, so that both matrices have zero-mean rows. For each matrix one can then define two possible covariance matrices:

\begin{align}
C_{u} & =\frac{1}{T-1}XX^{T}\Longrightarrow C_{u}\mathbf{u}_{i}=\lambda_{i}\mathbf{u}_{i}\\
C_{v} & =\frac{1}{N-1}X^{T}X\Longrightarrow C_{v}\mathbf{v}_{i}=\lambda_{i}\mathbf{v}_{i}
\end{align}
where $\mathbf{u}_{i}\in\mathbb{R}^{N\times1}$ and $\mathbf{v}_{i}\in\mathbb{R}^{T\times1}$ are the eigenvectors (singular vectors) associated with each of the above defined covariance matrices, which also share the same eigenvalues $\lambda_{i}$. In general, there are $r=\mathrm{rank}\left(X\right)$ non-zero eigenvectors of each type. The Singular Value Decomposition (SVD) theorem states that any observation matrix can be decomposed as:
\begin{align}
X & =USV^{T}\\
XV & =US\Longrightarrow X\mathbf{v}_{i}=\sigma_{i}\mathbf{u}_{i}
\end{align}
where $U\in\mathbb{R}^{N\times r}$ contains the $\mathbf{u}_{i}$
eigenvectors by columns, $V\in\mathbb{R}^{T\times r}$ contains the
$\mathbf{v}_{i}$ eigenvectors by columns, and $S\in\mathbb{R}^{r\times r}$
is a diagonal matrix containing the singular values $s_{ii}=\sigma_{i}=\sqrt{\lambda_{i}}$.

Assume one has performed SVD on the first trial observation matrix $X_{(1)}$, obtaining the singular vectors $U_{(1)}$ and $V_{(1)}$. 
Defining the projection matrix $P=U_{(1)}^{T}\in\mathbb{R}^{r\times N}$, which diagonalizes $C_{u}^{(1)}=\frac{1}{T-1}X_{(1)}X_{(1)}^{T}$, one can compute:interval = [0,1,3]

\begin{equation}
Y_{(1)}=PX_{(1)}\Longrightarrow\mathbf{y}_{i}^{(1)}=(\mathbf{u}_{i}^{(1)})^{T}X_{(1)}\in\mathbb{R}^{1\times T}
\end{equation}
where $Y_{(1)}\in\mathbb{R}^{r\times T}$ contains the projection of the first trial activity over its $r$ principal components. Now, the idea of the cv-PCA is to project on the same subspace the observations coming from the second trial \cite{Stringer-Nature}:
\begin{equation}
Y_{(2)}=PX_{(2)}\Longrightarrow\mathbf{y}_{i}^{(2)}=(\mathbf{u}_{i}^{(1)})^{T}X_{(2)}\in\mathbb{R}^{1\times T}
\end{equation}
and ask to form a covariance matrix with the product of both projections:
\begin{equation}
C_{\psi}=\dfrac{1}{T-1}Y_{(1)}Y_{(2)}^{T}=\dfrac{1}{T-1}\left(\begin{array}{ccc}
\mathbf{y}_{1}^{(1)}\cdot\mathbf{y}_{1}^{(2)} & \mathbf{y}_{1}^{(1)}\cdot\mathbf{y}_{2}^{(2)} & ...\\
\mathbf{y}_{2}^{(1)}\cdot\mathbf{y}_{1}^{(2)} & \mathbf{y}_{2}^{(1)}\cdot\mathbf{y}_{2}^{(2)} & ...\\
... & ... & \mathbf{y}_{r}^{(1)}\cdot\mathbf{y}_{r}^{(2)}
\end{array}\right).
\end{equation}
One can now hypothesize that the activity at time $t$ of any neuron $i$ during trial $k$ can be linearly decomposed as:
\begin{equation}
\mathbf{x}_{i}^{(k)}(t)=\mathbf{\psi}_{i}(t)+\mathbf{\epsilon}_{i}^{(k)}(t)\label{Decomposition}
\end{equation}
where $\mathbf{\psi}_{i}$ denotes the input-related activity, which should be independent of the trial for experiments carried in identical conditions; and $\mathbf{\epsilon}_{i}^{(k)}$ is defined as background or trial-to-trial variable activity, spanning a subspace that is orthogonal to the input-related one. One can then show  that the i-th diagonal element of $C_{\psi}$ is a non-biased estimator of the input-related variance $\omega_{i}$ \cite{Stringer-Nature}:
\begin{equation}
\omega_{i}=\dfrac{1}{T-1}\mathbf{y}_{i}^{(1)}\cdot\mathbf{y}_{i}^{(2)}=\dfrac{1}{T-1}\left((\mathbf{u}_{i}^{(1)})^{T}X_{(1)}\right)\left((\mathbf{u}_{i}^{(1)})^{T}X_{(2)}\right)^{T}=\dfrac{1}{T-1}(\mathbf{u}_{i}^{(1)})^{T}X_{(1)}X_{(2)}^{T}\mathbf{u}_{i}^{(1)}.
\end{equation}
Rewriting the observation matrices as $X_{(k)}=\Psi+\Sigma_{(k)}$,
where $\Psi$ contains vectors $\mathbf{\psi}_{i}$ in rows and
so does $\Sigma_{(k)}$ with vectors $\mathbf{\epsilon}_{i}^{(k)}$:
\begin{equation}
\omega_{i} =\dfrac{1}{T-1}\mathbf{u}_{i}^{T}\left(\Psi+\Sigma_{(1)}\right)\left(\Psi+\Sigma_{(2)}\right)^{T}\mathbf{u}_{i} =\dfrac{1}{T-1}\mathbf{u}_{i}^{T}\Psi\Psi^{T}\mathbf{u}_{i}+\dfrac{1}{T-1}\mathbf{u}_{i}^{T}\Sigma_{(1)}\Psi^{T}\mathbf{u}_{i}
\end{equation} 
where all the terms containing $\Sigma_{(2)}$ can be dismissed due to statistical independence of the realizations. Under reasonable approximations shown in \cite{Stringer-Nature}, one can prove that if the singular vectors $\mathbf{u}_{i}$ approach the singular vectors of $\Psi$, then the first term converges to the actual variance of the input-related activity along the principal direction $\mathbf{u}_{i}$, while the second term converges to zero. We have thus seen how the cv-PCA method proposed by Stringer \emph{et al.} allows one to estimate the input-related covariances. 

Is it possible to take a step further and find a proxy for the input-related activity observation matrix $\Psi\in\mathbb{R}^{N\times T}$, such that $\tilde{C}_{\psi}=\frac{1}{T-1}\Psi\Psi^{T}$ has the same eigenvalue spectrum as $C_{\psi}$? To do so, we first define a new basis of projected vectors given by:
\begin{equation}
\mathbf{z}_{i}=\sqrt{\mathbf{y}_{i}^{(1)}\cdot\mathbf{y}_{i}^{(2)}}\dfrac{\mathbf{y}_{i}^{(2)}}{\left\Vert \mathbf{y}_{i}^{(2)}\right\Vert }\in\mathbb{R}^{1\times T}.
\end{equation}
which clearly fulfills $\mathbb{E}\left[\mathbf{z}_{i}\cdot\mathbf{z}_{i}^{T}\right]=\omega_{i}$
under the same conditions as above. If $Z\in\mathbb{R}^{r\times T}$ is the matrix composed by the $r$ vectors $\mathbf{z}_{i}$ in rows, then:
\begin{equation}
\Psi=P^{T}Z.
\end{equation}
is the input-related activity of the neurons. From there, it is straightforward using  Eq.\ref{Decomposition} that one can estimate the background activity just by subtracting the input-related activity from the raw data:
\begin{equation}
\Sigma_{(k)}=X_{(k)}-\Psi.
\end{equation} 
Thus, following the above steps, we have found a proxy (Eq.\ref{Decomposition}) to decompose the overall activity of any neuron at a certain time step $t$ into an encoding (input-related) and background contributions. 

For the sake of consistency, Fig.\ref{spectra}) shows the
spectrum of the covariance matrix as estimated from such a decomposition, illustrating that, to a very good approximation it coincides with the above-computed one (using the standard cv-PCA method as discussed in \cite{Stringer-Nature}.

\newpage

\clearpage
\newpage

\clearpage
\newpage

 \clearpage \newpage 

 \clearpage \newpage 

 \clearpage \newpage 

 \clearpage \newpage

 \clearpage \newpage

 \clearpage \newpage 

 \clearpage \newpage 

 \clearpage \newpage 

 \clearpage \newpage 

 \clearpage \newpage

 \clearpage \newpage

\begin{figure}
\centering
\includegraphics[width=14cm]{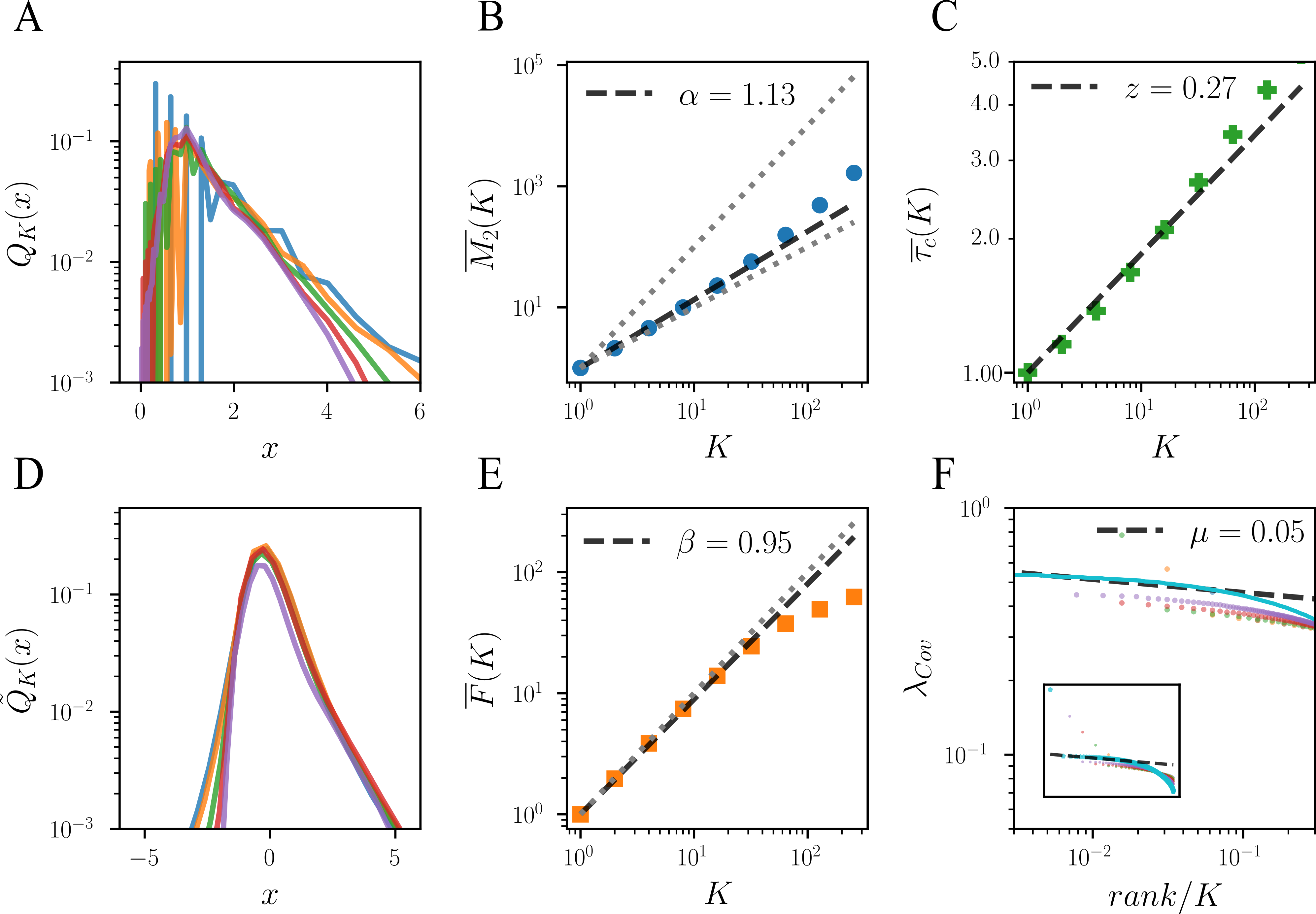}
\caption{RG analysis over surrogated data created by randomizing the spikes across neurons but not time. (A) Probability distribution for normalized non-zero activity in clusters of size $K=\{16, 32, 64, 128, 256\}$, corresponding to the yellow, blue, green, red and purple curves, respectively. (B) Scaling of activity variance inside clusters with cluster size. (C) Scaling of characteristic autocorrelation time with cluster size. (D) Probability distribution for normalized activity in momentum space, setting a cutoff on the first $K=\{16, 32, 64, 128, 256\}$ eigenmodes (computed as in \cite{Bialek-largo}, color code as in panel A) (E) Scaling of the "free-energy" with cluster size. (F) Inset: Covariance matrix spectrum with (almost) flat power law fit against the normalized rank and strong first eigenvalue outlayer. Main: closer view}
\label{shuffledneurons}
\end{figure} \clearpage \newpage

\begin{figure}
\centering
\includegraphics[width=14cm]{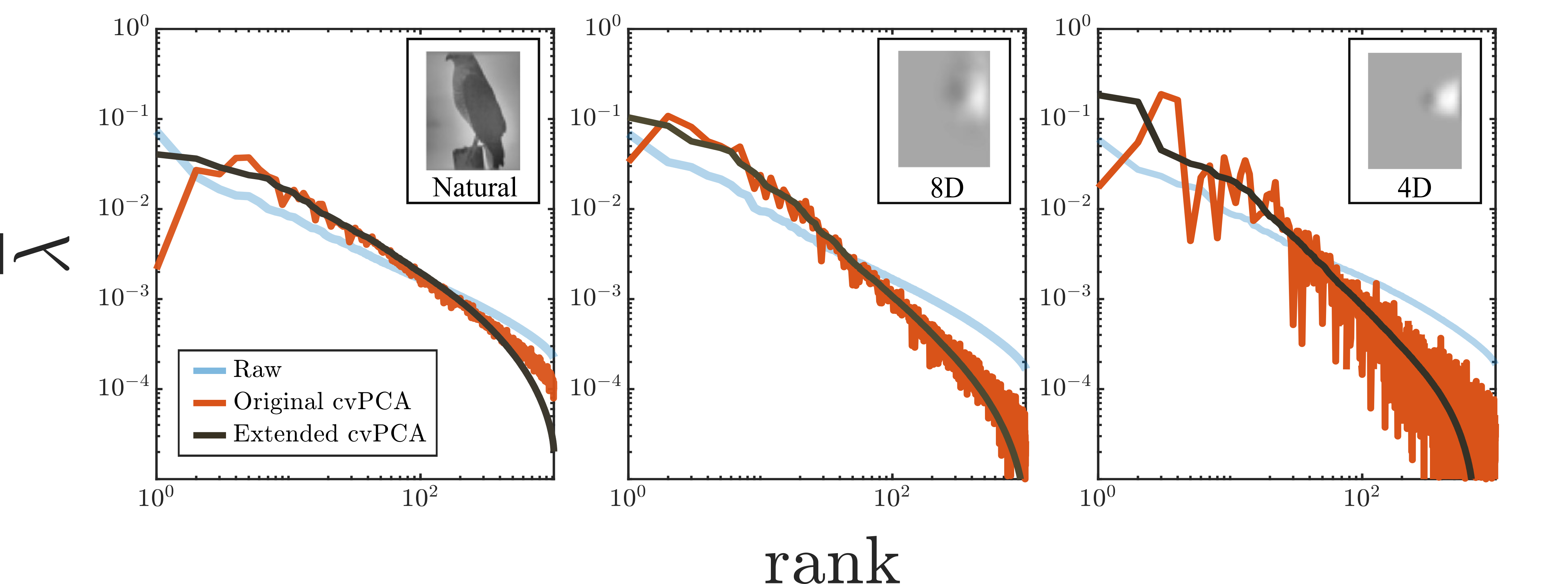}
\caption{Distribution of pair-wise covariance eigenvalues (normalized as in \cite{Stringer-Nature} by their sum) measured from the activity of mice subject to three different types of stimuli: natural images (left panel), and their eight-dimensional (middle panel) and four-dimensional (right panel) projections. For each panel, the blue curve marks the spectrum measured on the overall, raw activity. Orange curves correspond to the spectrum of the input-related or encoding activity as measured by the cvPCA method proposed in \cite{Stringer-Nature}. On the other hand, black curves also represent the input-related covariance spectrum, but they are extracted directly from the proposed proxy for the input-related activity, as described in Section 3 of this SI.}\label{spectra}
\end{figure} \clearpage \newpage 

\begin{table}
\centering
\setlength{\tabcolsep}{3pt}
\caption{\label{tab:Table1}For each region we show: \textbf{(i)} the number $M$ of experiments
considered; \textbf{(ii)} the average number of neurons across experiments;
\textbf{(iii)} the average length of the recordings across experiments
(once all intervals of spontaneous activity in an experiment have
been merged together); \textbf{(iv)} the average lifetime and standard
deviation across experiments of the correlations exponential decay,
as given by the average autocorrelation function across all neurons
in a region; \textbf{(v)} the average percentage of non-stationary
neurons in a region; \textbf{(vi)} the distance to the edge of instability, as inferred in \cite{Helias}, with the standard deviation across experiments; \textbf{(vii)} same as in (vi), but using the method proposed in \cite{YuHu}; \textbf{(viii)} the error using the Cramer-von Mises statistic between the sampled eigenvalue distribution and the theoretical distribution proposed in \cite{YuHu} for a model of linear rate neurons with recurrent connectivity (RC); \textbf{(ix)} same as in (viii), but comparing the empirical distribution to a best-fitting Marchenko-Pastur (MP) distribution; \textbf{(x)} estimated distance to the edge of instability using a common extrapolated size of $N=10^4$ neurons, with the error given as the standard deviation across experiments. For a given region, each experiment corresponds to a different mouse, and only mice having more than $N=128$ recorded were selected.}
\vspace{0.7cm}
\resizebox{\linewidth}{!}{
\begin{tabular}{|l|l|c|c|c|c|c|c|c|c|c|c|}
\hline 
\multirow{2}{*}{{\small{}Full name}} & \multirow{2}{*}{{\small{}Abbrev.}} & \multirow{2}{*}{$M$} & \multirow{2}{*}{$N\left(\times10^{2}\right)$} & \multirow{2}{*}{$T\left(\times10^{2}s\right)$} & \multirow{2}{*}{$\tau_{Corr}(ms)$} & \multirow{2}{*}{\% N.S.} & \multirow{2}{*}{{\small{}$\lambda_{max}$}} & \multirow{2}{*}{$g$} & \multirow{2}{*}{$L_{RC}^{2}$} & \multirow{2}{*}{$L_{MP}^{2}$} & \multirow{2}{*}{{\small{}$\lambda_{max}^{N=1e4}$}}\tabularnewline
 &  &  &  &  &  &  &  &  &  &  & \tabularnewline
\hline 
{\small{}Anterior cingulate area} & {\small{}ACA} & $3$ & $1.8\pm0.8$ & $7.5\pm3.2$ & $0.10\pm0.02$ & $1.2\pm0.5$ & $0.70\pm0.09$ & $0.74\pm0.09$ & $0.07\pm0.02$ & $0.11\pm0.04$ & $0.965\pm0.013$\tabularnewline
{\small{}Basolateral amygdalar nucleus} & {\small{}BLA} & $2$ & $2.7\pm0.0$ & $8.1\pm0.5$ & $0.09\pm0.01$ & $1.3\pm0.6$ & $0.81\pm0.00$ & $0.86\pm0.04$ & $0.07\pm0.02$ & $0.17\pm0.02$ & $0.969\pm0.000$\tabularnewline
{\small{}Cornu ammonis} & {\small{}CA1} & $2$ & $1.9\pm0.2$ & $8.8\pm2.9$ & $0.07\pm0.01$ & $7.7\pm0.3$ & $0.85\pm0.04$ & $0.83\pm0.05$ & $0.06\pm0.01$ & $0.15\pm0.02$ & $0.980\pm0.004$\tabularnewline
{\small{}Caudoputamen} & {\small{}CP} & $4$ & $4.1\pm1.0$ & $11.0\pm1.2$ & $0.09\pm0.01$ & $2.6\pm1.4$ & $0.91\pm0.02$ & $0.93\pm0.01$ & $0.09\pm0.02$ & $0.23\pm0.01$ & $0.983\pm0.002$\tabularnewline
{\small{}Lateral septal nucleus} & {\small{}LS} & $3$ & $2.6\pm0.5$ & $6.4\pm2.9$ & $0.05\pm0.03$ & $2.1\pm0.5$ & $0.90\pm0.02$ & $0.83\pm0.02$ & $0.05\pm0.00$ & $0.15\pm0.00$ & $0.972\pm0.007$\tabularnewline
Dorsal part of the lateral geniculate complex & LGd & $5$ & $1.6\pm0.1$ & $8.3\pm1.0$ & $0.08\pm0.01$ & $1.6\pm1.6$ & $0.77\pm0.04$ & $0.82\pm0.06$ & $0.09\pm0.02$ & $0.12\pm0.03$ & $0.962\pm0.009$\tabularnewline
Lateral posterior nucleus of the thalamus & LP & $2$ & $1.8\pm0.1$ & $9.4\pm0.7$ & $0.08\pm0.01$ & $0.3\pm0.3$ & $0.72\pm0.06$ & $0.66\pm0.06$ & $0.07\pm0.04$ & $0.06\pm0.00$ & $0.983\pm0.002$\tabularnewline
Medial geniculate complex of the thalamus & MG & $2$ & $2.2\pm0.2$ & $10.0\pm2.4$ & $0.09\pm0.01$ & $3.4\pm3.4$ & $0.74\pm0.07$ & $0.73\pm0.00$ & $0.09\pm0.02$ & $0.08\pm0.01$ & $0.962\pm0.009$\tabularnewline
{\small{}Primary motor area} & {\small{}MOp} & $3$ & $4.2\pm2.2$ & $11.5\pm0.9$ & $0.12\pm0.02$ & $1.3\pm0.5$ & $0.93\pm0.01$ & $0.91\pm0.02$ & $0.09\pm0.01$ & $0.27\pm0.02$ & $0.987\pm0.002$\tabularnewline
{\small{}Secondary motor area} & {\small{}MOs} & $5$ & $2.5\pm0.7$ & $8.5\pm3.2$ & $0.10\pm0.01$ & $4.3\pm1.8$ & $0.84\pm0.02$ & $0.84\pm0.03$ & $0.06\pm0.01$ & $0.16\pm0.02$ & $0.975\pm0.000$\tabularnewline
{\small{}Orbital area} & {\small{}ORB} & $3$ & $2.7\pm0.3$ & $9.5\pm0.8$ & $0.09\pm0.02$ & $1.8\pm0.7$ & $0.87\pm0.02$ & $0.80\pm0.02$ & $0.05\pm0.01$ & $0.13\pm0.01$ & $0.980\pm0.003$\tabularnewline
{\small{}Prelimbic area} & {\small{}PL} & $4$ & $2.4\pm0.5$ & $8.7\pm1.6$ & $0.12\pm0.04$ & $3.9\pm2.3$ & $0.83\pm0.05$ & $0.80\pm0.08$ & $0.04\pm0.01$ & $0.14\pm0.05$ & $0.974\pm0.010$\tabularnewline
{\small{}Posterior complex of the thalamus} & {\small{}PO} & $3$ & $2.8\pm1.1$ & $10.1\pm1.0$ & $0.10\pm0.03$ & $0.9\pm0.6$ & $0.89\pm0.02$ & $0.78\pm0.05$ & $0.09\pm0.02$ & $0.12\pm0.03$ & $0.983\pm0.005$\tabularnewline
{\small{}Superior culliculus, intermediate gray layer} & {\small{}SCig} & $2$ & $2.8\pm0.9$ & $8.4\pm0.4$ & $0.11\pm0.00$ & $3.6\pm2.0$ & $0.91\pm0.01$ & $0.86\pm0.05$ & $0.08\pm0.02$ & $0.19\pm0.04$ & $0.986\pm0.001$\tabularnewline
{\small{}Primary somatosensory area} & {\small{}SSp} & $2$ & $3.3\pm0.0$ & $11.6\pm1.0$ & $0.12\pm0.01$ & $1.7\pm0.7$ & $0.93\pm0.01$ & $0.91\pm0.02$ & $0.08\pm0.01$ & $0.20\pm0.02$ & $0.988\pm0.002$\tabularnewline
Primary visual area & VISp & $4$ & $1.9\pm0.3$ & $9.5\pm2.1$ & $0.08\pm0.01$ & $4.0+2.5$ & $0.84\pm0.03$ & $0.85\pm0.02$ & $0.07\pm0.00$ & $0.16\pm0.02$ & $0.978\pm0.004$\tabularnewline
\hline 
\end{tabular}
}
\end{table}
\clearpage
\newpage

\begin{table}
\centering
\setlength{\tabcolsep}{3pt}
\caption{\label{tab:Table2}For each region we show the average bin size $\Delta t$ used in the RG analysis (with standard deviation across experiments) as given
by the geometric mean of the ISIS distribution. Moreover, for each
region and measured exponents we collect: \textbf{(i)} the average
value across experiments; \textbf{(ii)} the mean-absolute-error, computed
as the average across experiments of the experiment-specific errors
measured over split-quarters of data; \textbf{(iii)} the standard
deviation across experiments; \textbf{(iv)} the R-squared value of
the best powerlaw fit for the experiment with a greater number of
recorded neurons. For the exponent of the rank-ordered covariance
matrix eigenvalues, we also provide the R-squared value of the best
exponential fit, as well as the likelihood ratios and p-values for
the significance of each test comparing the powerlaw fit of the associated eigenvalue density with corresponding
exponential and lognormal distributions (statistically significant
p-values are highlighted in boldface). Positive ratios indicate that
the data is best fitted by a powerlaw distribution. For a given region,
each experiment corresponds to a different mouse, and only mice having
more than $N=128$ recorded neurons were selected.}
\vspace{0.7cm}
\resizebox{\linewidth}{!}{
\begin{tabular}{|c|c|cccc|cccc|cccc|ccccccccc|}
\hline 
\multirow{2}{*}{Abbrev.} & \multirow{2}{*}{{\small{}$\Delta t(s)$}} & \multirow{2}{*}{$\left\langle \boldsymbol{\alpha}\right\rangle $} & \multirow{2}{*}{MAE} & \multirow{2}{*}{$\sigma$} & \multirow{2}{*}{$r^{2}$} & \multirow{2}{*}{$\left\langle \boldsymbol{\beta}\right\rangle $} & \multirow{2}{*}{MAE} & \multirow{2}{*}{$\sigma$} & \multirow{2}{*}{$r^{2}$} & \multirow{2}{*}{$\left\langle \mathbf{z}\right\rangle $} & \multirow{2}{*}{MAE} & \multirow{2}{*}{$\sigma$} & \multirow{2}{*}{$r^{2}$} & \multirow{2}{*}{$\left\langle \boldsymbol{\mu}\right\rangle $} & \multirow{2}{*}{MAE} & \multirow{2}{*}{$\sigma$} & \multirow{2}{*}{$r_{\mathrm{pow}}^{2}$} & \multirow{2}{*}{$r_{\mathrm{exp}}^{2}$} & \multicolumn{2}{c}{Exponential} & \multicolumn{2}{c|}{Lognormal}\tabularnewline
 &  &  &  &  &  &  &  &  &  &  &  &  &  &  &  &  &  &  & LR & $p$ & LR & $p$\tabularnewline
\hline 
ACA & $0.16\pm0.07$ & 1.43 & 0.02 & 0.11 & 1.00 & 0.81 & 0.03 & 0.05 & 1.00 & 0.2 & 0.02 & 0.06 & 0.98 & 0.74 & 0.05 & 0.16 & 0.83 & 0.58 & 90.5 & \textbf{0.00} & -0.1 & 0.22\tabularnewline
BLA & $0.13\pm0.01$ & 1.25 & 0.03 & 0.02 & 1.00 & 0.87 & 0.04 & 0.03 & 1.00 & 0.21 & 0.03 & 0.03 & 1.00 & 0.82 & 0.03 & 0.00 & 0.85 & 0.78 & 9.4 & 0.12 & -2.9 & 0.09\tabularnewline
CA1 & $0.13\pm0.04$ & 1.37 & 0.03 & 0.02 & 1.00 & 0.78 & 0.04 & 0.05 & 1.00 & 0.18 & 0.03 & 0.01 & 0.99 & 0.78 & 0.08 & 0.02 & 0.89 & 0.65 & 55.4 & \textbf{0.00} & -0.3 & 0.59\tabularnewline
CP & $0.08\pm0.02$ & 1.38 & 0.02 & 0.12 & 1.00 & 0.77 & 0.05 & 0.06 & 0.98 & 0.16 & 0.02 & 0.03 & 0.99 & 0.98 & 0.1 & 0.15 & 0.96 & 0.67 & 43.3 & \textbf{0.00} & -2.5 & 0.17\tabularnewline
LS & $0.17\pm0.13$ & 1.41 & 0.05 & 0.08 & 1.00 & 0.81 & 0.04 & 0.07 & 1.00 & 0.25 & 0.05 & 0.08 & 0.99 & 0.86 & 0.08 & 0.15 & 0.94 & 0.86 & 8.4 & \textbf{0.05} & -0.9 & 0.32\tabularnewline
LGd & $0.07\pm0.03$ & 1.26 & 0.03 & 0.06 & 1.00 & 0.82 & 0.04 & 0.06 & 1.00 & 0.24 & 0.06 & 0.03 & 1.00 & 0.67 & 0.03 & 0.06 & 0.9 & 0.9 & 0.5 & 0.86 & -1.5 & 0.21\tabularnewline
LP & $0.08\pm0.02$ & 1.36 & 0.03 & 0.07 & 1.00 & 0.77 & 0.02 & 0.02 & 1.00 & 0.23 & 0.05 & 0.06 & 0.99 & 0.52 & 0.06 & 0.06 & 0.99 & 0.63 & 152.3 & \textbf{0.00} & -3.5 & 0.08\tabularnewline
MG & $0.06\pm0.01$ & 1.23 & 0.02 & 0.05 & 1.00 & 0.86 & 0.02 & 0.04 & 1.00 & 0.18 & 0.04 & 0.05 & 0.99 & 0.55 & 0.05 & 0.00 & 0.89 & 0.75 & 24.3 & \textbf{0.00} & -0.0 & 0.90\tabularnewline
MOp & $0.06\pm0.01$ & 1.5 & 0.03 & 0.01 & 1.00 & 0.78 & 0.03 & 0.06 & 0.99 & 0.22 & 0.02 & 0.05 & 1.00 & 0.92 & 0.05 & 0.05 & 0.85 & 0.45 & 138.8 & \textbf{0.00} & -0.2 & 0.66\tabularnewline
MOs & $0.16\pm0.03$ & 1.39 & 0.04 & 0.03 & 1.00 & 0.81 & 0.04 & 0.04 & 1.00 & 0.22 & 0.03 & 0.06 & 0.99 & 0.85 & 0.07 & 0.04 & 0.93 & 0.56 & 85.1 & \textbf{0.00} & -0.0 & 0.87\tabularnewline
ORB & $0.12\pm0.02$ & 1.43 & 0.02 & 0.02 & 1.00 & 0.74 & 0.02 & 0.03 & 0.99 & 0.16 & 0.02 & 0.03 & 0.99 & 0.77 & 0.03 & 0.03 & 0.88 & 0.49 & 146.5 & \textbf{0.00} & -0.6 & 0.50\tabularnewline
PL & $0.13\pm0.08$ & 1.42 & 0.04 & 0.10 & 1.00 & 0.82 & 0.04 & 0.1 & 0.99 & 0.19 & 0.03 & 0.02 & 0.99 & 0.77 & 0.05 & 0.13 & 0.86 & 0.52 & 180.9 & \textbf{0.00} & -0.6 & 0.48\tabularnewline
PO & $0.07\pm0.02$ & 1.43 & 0.02 & 0.04 & 1.00 & 0.74 & 0.03 & 0.04 & 1.00 & 0.32 & 0.01 & 0.02 & 0.99 & 0.62 & 0.03 & 0.09 & 0.91 & 0.76 & 22.8 & \textbf{0.00} & 0.0 & 0.85\tabularnewline
SCig & $0.07\pm0.00$ & 1.31 & 0.04 & 0.01 & 1.00 & 0.82 & 0.04 & 0.01 & 0.98 & 0.21 & 0.04 & 0.02 & 1.00 & 0.77 & 0.08 & 0.06 & 0.96 & 0.76 & 26.6 & \textbf{0.00} & -2.8 & 0.11\tabularnewline
SSp & $0.06\pm0.01$ & 1.51 & 0.04 & 0.01 & 1.00 & 0.77 & 0.02 & 0.12 & 0.99 & 0.27 & 0.02 & 0.01 & 1.00 & 0.91 & 0.03 & 0.04 & 0.82 & 0.44 & 121.4 & \textbf{0.00} & -0.5 & 0.54\tabularnewline
VISp & $0.12\pm0.04$ & 1.42 & 0.02 & 0.06 & 1.00 & 0.73 & 0.02 & 0.05 & 1.00 & 0.15 & 0.02 & 0.02 & 0.98 & 0.90 & 0.04 & 0.14 & 0.98 & 0.75 & 22.2 & \textbf{0.00} & -1.2 & 0.26\tabularnewline
\hline 
\end{tabular}
}
\end{table}

\clearpage
\newpage

\bibliographystyle{unsrt}
\bibliography{pnas-sample}